\documentclass[12pt,aps,pra,eqsecnum,psfig,amssymb,subeqn,graphicx,colordvi]{article}
\usepackage{amssymb}
\usepackage{color,times,epsfig}
\newcommand{\eqref}[1]{(\ref{#1})}
\newcommand{\be}{\begin{equation}}
\newcommand{\ee}{\end{equation}}
\newcommand{\ba}{\begin{array}}
\newcommand{\ea}{\end{array}}
\newcommand{\beqa}{\begin{eqnarray}}
\newcommand{\eeqa}{\end{eqnarray}}

\newtheorem{defi}{Definition}[section]
\newcommand{\sdir}{\ensuremath{\rlap{\raisebox{.15ex}{$ \mskip 6.0
mu\scriptstyle+ $ }}\supset}}

\begin{document}




\title{ \mathversion{bold}   $sh(2/2) $ Superalgebra
eigenstates and  generalized supercoherent and supersqueezed
states}

\author{Nibaldo Alvarez--Moraga,${}^{1,2,3}$  and V\'eronique Hussin${}^{1,2,4}$}
\maketitle

\footnotetext[1]{Centre de Recherches Math\'ematiques,
Universit\'e de Montr\'eal, C.P. 6128, Succ.~Centre-ville,
Montr\'eal (Qu\'ebec) H3C 3J7, Canada.}
\footnotetext[2]{D\'epartement de Math\'ematiques et de
Statistique, Universit\'e de Montr\'eal, C.P. 6128,
Succ.~Centre-ville, Montr\'eal (Qu\'ebec) H3C 3J7, Canada.}
\footnotetext[3]{Electronic mail address:
alvarez@dms.umontreal.ca}\footnotetext[4]{Electronic mail address:
hussin@dms.umontreal.ca}

\begin{abstract} The superalgebra eigenstates (SAES) concept is
introduced and then applied to find the SAES associated to the
$sh(2/2)$ superalgebra, also known as Heisenberg--Weyl Lie
superalgebra. This implies to solve a Grassmannian eigenvalue
superequation. Thus,  the $sh(2/2)$ SAES contain the class of
supercoherent states associated to the supersymmetric harmonic
oscillator and also a class of supersqueezed states associated to
the $osp(2/2) \sdir sh(2/2)$ superalgebra, where $osp(2/2)$
denotes the orthosymplectic Lie superalgebra generated by the set
of operators formed from the quadratic products of the
Heisenberg--Weyl Lie superalgebra generators. The properties of
these states are investigated and compared with those of the
states obtained by applying the group-theoretical technics.
Moreover, new classes of generalized supercoherent and
supersqueezed states are also obtained. As an application, the
superHermitian and $\eta$--pseudo--superHermitian  Hamiltonians
without a defined Grassmann parity and isospectral to the harmonic
oscillator are constructed. Their eigenstates and associated
supercoherent states are calculated.
\end{abstract}

\baselineskip 0.75cm

\newpage

\section{Introduction} \label{sec-dorila}

The algebra eigenstates (AES) associated to a real Lie algebra
have been defined  as the set of eigenstates  of an arbitrary
complex linear combination of the generators of the considered
algebra\cite{Brif0,Brif}. According to the particular realization
of the Lie algebra generators, the determination of the AES
implies, for instance, to solve an ordinary or a partial
differential equation, to apply the operator technics, etc. For
example, in the case of the  $su(2)$ Lie algebra, different
approaches have been used such as the constellation
formalism\cite{Ba}, the ordinary first order differential
equations\cite{Brif} or the operator method\cite{Nam}. The same
methods have also been applied to find the AES for the $su(1,1)$
Lie algebra\cite{Nam,Brif}. In the case of the two-photon AES,
associated to the $su(1,1) \sdir h(2)$ Lie algebra,  used have
been done of ordinary second order differential
equation\cite{Brif0}. More recently,  the AES associated to the
$h(2) \oplus su(2)$ Lie algebra have been obtained using these
types of methods\cite{NaVh}. In particular\cite{Brif} it has
been demonstrated that the generalized coherent states (GCS)
associated to the  $SU(2)$ and $SU(1,1)$ Lie groups, based on
group-theoretical approach\cite{Pere}, are subsets of the sets of
AES  associated to their corresponding Lie algebras. Moreover, the
super coherent states of the supersymmetric harmonic
oscillator\cite{JwAsSaVh} as defined by Aragone and
Zypmann\cite{ArZy} and a new class of supercoherent and
supersqueezed states regarded as minimum uncertainty states have
been obtained\cite{NaVh}. Generalized supercoherent states (GSCS)
associated to  Lie supergroups have also been calculated
following a generalized  group-theoretical approach. This is the
case, for example,  of the supercoherent states associated to the
following supergroups: Heisenberg--Weyl (H--W) and
$OSp(1/2)$\cite{Faetal}, $U(1/2)$\cite{HuLMNi,Sa},
$U(1/1)$\cite{ApCt} and  $OSp(2/2) $ \cite{AmElGraLMMi}.

In the view  of these approaches we ask the question of how  we
can generalize the AES concept valid for Lie algebras to Lie
superalgebras. In general, as the even subspace of a Lie
superalgebra is an ordinary  Lie algebra, it is clair that the new
concept must generalize in an appropriate form the AES concept.
Indeed, the set of superalgebra eigenstates (SAES) associated to
linear combinations of even generators of the Lie superalgebra
must contain the AES associated to the Lie algebra generated by
these generators. Moreover, we expect that the SAES associated to
a certain class of superalgebras contain the  GSCS  of the related
Lie supergroups. Another criterion to define the SAES concept
start from the utility  that we can give to this concept when we
study a particular quantum system, more precisely when we want to
know the eigenstates of a physical observable represented by a
superHermitian operator formed by a  linear combination of the
superalgebra generators or by a suitable  product of these
generators. According with these requirements, we propose  the
following definition of the SAES concept.

\begin{defi}  The SAES associated to a Lie superalgebra correspond
to the set of eigenstates  of an arbitrary linear combination,
with coefficients in the Grassmann algebra ${\mathbb C} B_L, $ of
the superalgebra generators. This means that if $\cal L $ is a
superalgebra  generated by the set of even operators $ \Phi( a_1)
, \Phi( a_2 ) , \ldots \Phi (a_m ) $ and the set of odd operators
$ \Phi ( a_{m+1}), \Phi ( a_{m+2} ),   \ldots, \Phi ( a_{m+n}),$
the SAES associated to $\cal L $ are determined by the eigenvalue
equation  \be \left[ \sum_{i=1}^{m+n} B^i \Phi( a_i) \right] |\psi
\rangle = Z | \psi \rangle, \label{gen-super-equation}\ee where
$B^i \in {\mathbb C} B_L, \;  \forall i=1,2, \ldots , m+n $ and $Z
\in {\mathbb C} B_L . $
\end{defi}
In general, the superstate $ |\psi \rangle $ is a linear
combination, with coefficients in ${\mathbb C} B_L, $ of the basis
vectors of a graded superHilbert space ${\cal W},$ the representation
space of the superalgebra on which it acts.

Let us here mention that the Appendix \ref{sec-Not} contains the
notations and conventions used in the context of Grassmann
algebras, Lie superalgebras and supergroups. This will help for a
good understanding of this work.

From the preceding definition,  we see that to  know explicitly
the SAES associated to a given Lie superalgebra, we must analyze
case by case the different possible solutions  of the Grassmannian
eigenvalue equation \eqref{gen-super-equation} taking into account
both the domain of definition of the Grassmann coefficients and
the parity of them. In general, the calculations can  be long and
fastidious, but in physical applications, some simplifications
appear due to some constrains on the coefficients like assuming a
certain type of parity.

A natural generalization of the concept of AES to SAES starts with
H--W superalgebra $sh(2/2)$ generated by the bosonic operators $a,
a^\dagger$ and $I$ and the fermionic ones $b$ and $b^\dagger .$ We
expect to recover the usual algebra
eigenstates\cite{NaVh,ArZy,OrSa} but also  supercoherent and
supersqueezed states based on a  group theoretical
approach\cite{KoNiTr,MMNieto}.

Let us remind that the well-known bosonic algebra is generated by
the even operators $a, a^\dagger$ and $I,$ that satisfy the usual
non-zero commutation relation \be [a ,a^\dagger] =I,
\label{com-aad}\ee and act on the usual Fock space $ {\cal F}_b =
\{ |n \rangle, \ n \in {\mathbb N}\}, $ as follows \be a |n
\rangle = \sqrt{n} | n-1 \rangle, \qquad a^\dagger |n \rangle =
\sqrt{n+1} | n+1 \rangle, \quad n \in {\mathbb N} .
\label{raising-a} \ee The operators $a, a^\dagger$ are the usual
annihilation and creation operators of the harmonic oscillator,
and $I$ acts as the identity operator. The corresponding fermionic
superalgebra is generated by the odd operators $b, b^\dagger$ and
the even operator $I,$  which satisfy the non-zero super
commutation relation \be  \{ b, b^\dagger \} =I.
\label{supercom-bbd}\ee These operators act on the graded space $
{\cal F}_f = \{  |+ \rangle, |- \rangle \}$ as follows \be
b|+\rangle = |- \rangle, \quad b | - \rangle =0, \quad  b^\dagger
|+\rangle = 0, \quad b^\dagger | - \rangle = |+\rangle.
\label{raising-b} \ee Taking the all set $\{a, a^\dagger , I, b ,
b^\dagger \}$ satisfying the non-zero supercommutation relations
\eqref{com-aad} and \eqref{supercom-bbd}, we get the H--W
superalgebra $sh(2/2).$ Its acts naturally on the graded Fock
space ${\cal F}_b \otimes {\cal F}_f = \{ | n , \pm\rangle, \ n
\in {\mathbb N} \}.$ In order to compute the SAES of this
superalgebra we will consider linear combinations  over the field
of Grassmann numbers. This means that, in general, we will deal
with linear combinations of the bosonic (even) and fermionic (odd)
operators with the coefficients taking values in the set ${\mathbb
C}B_L.$

The paper will be thus distributed as follows. In section
\ref{sec-hwla}, we will determine the SAES associated to the
bosonic H--W Lie algebra. A significant difference with respect to
the other approaches is now that linear combinations of generators
is considered over the field of Grassmann numbers. Connections
with preceding approaches will be made. In section
\ref{sec-fersuper}, fermionic H--W Lie superalgebra will be
considered. These special SAES  cases will give a good
understanding of the specificities induced by working with
Grassmann valued variables and will help us to give a complete
description of the SAES associated to the H--W Lie superalgebra in
section \ref{sec-super-hw-lie-super}. Finally, in section
\ref{sec-isospectral},  Hamiltonians which are isospectral to the
harmonic oscillator one  will be constructed and their
associated supercoherent states will be described. The notations
and conventions used in this work will be revised in the Appendix
\ref{sec-Not} whereas the details of calculus of the SAES of
section \ref{sec-super-hw-lie-super} will be presented in the
Appendix \ref{sec-caso-general}.

\section{SAES  associated to the   Heisenberg--Weyl Lie algebra,
generalized supercoherent and supersqueezed
states}\label{sec-hwla}

The SAES associated to the H--W  Lie algebra will be obtained as
the states $|\psi \rangle $ that verify the eigenvalue equation
\be \label{superH-Wtotal} [A_- a + A_+ a^\dagger + A_3 I ] | \psi
\rangle = Z | \psi \rangle, \ee  where $A_\pm , A_3$ and $Z  \in
{\mathbb C} B_L. $ From the structure of this equation, we expect
to recover the usual results concerning, in particular, the
eigenstates of $a,$ i.e., the  standard coherent states of the
harmonic oscillator\cite{Pere}. That is the reason why we begin
our considerations by taking first $A_+ = A_3 = 0 .$ In this
context, we will distinguish between  the cases where ${(A_-
)}_\phi $ is zero and not zero. Next, the general combination
\eqref{superH-Wtotal} will be considered with ${(A_- )}_\phi \ne 0
.$ This means that $A_- $ is an invertible Grassmann number and
the relation \eqref{superH-Wtotal} thus reduces to
 \be \label{eq-reduite} [a + \beta a^\dagger ] \psi \rangle = z | \psi \rangle,  \qquad
 \beta , \ z \ \in {\mathbb C}B_L .\ee

\subsection{Generalized coherent states}
If we take  $A_+ = A_3 =0, $ the eigenvalue equation
\eqref{superH-Wtotal} thus writes \be \label{eigen-ann} A_- \ a |
\psi \rangle = Z | \psi \rangle. \ee Let us assume  a solution of
the type \be \label{fon-serie-n} | \psi \rangle = \sum_{n=0}^\infty
C_n |n \rangle, \qquad C_n \ \in {\mathbb C } B_L . \ee By inserting
(\ref{fon-serie-n}) in (\ref{eigen-ann}), applying (\ref{raising-a})
and using the orthogonality property of states ${ \{  | n \rangle \}
}_{n=0}^{\infty}, $ we get to the following recurrence relation \be
\label{rec-one} A_-   C_{n+1} = {Z C_n \over \sqrt{n+1}}, \qquad
n=0,1, \ldots. \ee Here we must consider two cases: the cases ${(A_-
)}_\phi \ne 0$ and $ {(A_- )}_\phi = 0. $

In the first  case,  ${(A_- )}_\phi \ne 0 $ is thus an invertible
quantity  and we can isolate the coefficient $ C_{n+1} $  in
\eqref{rec-one}. It is easy to show that we get: \be C_n =
{{\left( ({A_- )}^{-1} Z \right)}^n \over \sqrt{n!}} C_0 , \qquad
n=1,2,\ldots.\ee The SAES associated to the operator $ A_- a $
with eigenvalue $Z$ are then given by \be | z \rangle =
\sum_{n=0}^{\infty} {z^n \over \sqrt{n!}} C_0 | n \rangle =
\sum_{n=0}^{\infty} { {(z a^\dagger )}^n  \over n! } C_0 | 0
\rangle= e^{z a^\dagger} C_0 | 0 \rangle, \ee where $z={(A_-
)}^{-1} Z . $   As we are interested in normalized eigenstates, we
take $ (C_0)_{\phi} \ne 0 $ and the  eigenstates can be written as
\be  | z \rangle = {\mathbb D} (z_0 ) {\mathbb D} (z_1) | 0
\rangle, \label{gen-cohe-states} \ee where \be
 {\mathbb D}
(z_0 ) = \exp\left( z_0 a^\dagger - z_0^\ddagger a \right), \qquad
{\mathbb D} (z_1 ) = \exp\left( z_1 a^\dagger - z_1^\ddagger a
\right), \label{mathd-super-adj} \ee  $ z_0 =   {\left( ({A_-
)}^{-1} Z \right)}_0 $ and $ z_1 = {\left( ({A_- )}^{-1} Z
\right)}_1 .$

We notice that the generalized coherent states associated to the
harmonic oscillator system, considered as eigenstates of the
annihilation operator $a,$ are here given by
(\ref{gen-cohe-states}) when $ A_- = \epsilon_\phi, $ i.e., when
$z_0 = Z_0 $ and $z_1=Z_1.$ This states are obtained  by  applying
successively the superunitary operators  ${\mathbb D} (Z_1 )$ and
${\mathbb D} (Z_0 )$ to the fundamental state $|0\rangle .$

In the second case,  that is when ${(A_- )}_\phi = 0, $ we can not
obtain a simple closed expression  to describe all the algebra
eigenstates.  A class of solution is:  \be C_n = { {\left( C_1 {(
C_0 )}^{-1} \right )}^n \over \sqrt{n!} } C_0 , \qquad
n=2,3,\ldots, \ee together with \be A_- C_1 = Z C_0
\label{cond-uno} \ee and $C_0 $ is an arbitrary coefficient such
that $ {(C_0)}_\phi \ne 0 . $ The condition (\ref{cond-uno})
implies that $Z_\phi = 0 $ and is equivalent to the following
system of superequations \beqa
{( A_- )}_0 { (C_1 )}_0 +  {( A_- )}_1 { (C_1 )}_1 = Z_0  { (C_0 )}_0 +  Z_1 { (C_0 )}_1, \\
{( A_- )}_0 { (C_1 )}_1 +  {( A_- )}_1 { (C_1 )}_0 = Z_0 { (C_0
)}_1 +  Z_1 { (C_0 )}_0, \eeqa where we have decomposed $A_- , C_0
, C_1 $ and $Z$ into their even and odd parts. This system can be
solved to give $ C_1 $ in terms of $C_0 . $ A set of normalized
eigenstates corresponding to the eigenvalue $Z= \alpha  A_-, \
\alpha \ \in {\mathbb C},$ is given by the standard coherent
states \be |\alpha  \epsilon_{\phi} \rangle = \exp\left( \alpha
\epsilon_\phi  a^\dagger -  {\bar \alpha} \epsilon_\phi a \right)
|0\rangle = {\mathbb D}(\alpha \epsilon_\phi ) |0 \rangle .\ee So,
in the special case when ${(A_-)}_0 = 0, $ the algebra eigenstates
of the odd operator $ {(A_-)}_1 a $ contain the set of coherent
states of the standard harmonic oscillator.

\subsubsection{Density of algebra}
It is interesting to mention that we can interpret this last
result in terms of the concept of density of algebra. Indeed, let
us define the odd operators \be \label{dens-linear} {\mathbb A}_-
= z_1^\ddagger a, \qquad  {\mathbb A}_+ = - z_1 a, \quad z_1 \ \in
{\mathbb C}B_{L_1} . \ee By integrating these operators with
respect to the corresponding odd variable, we get \be
\label{den-anni-crea} a = \int {\mathbb A}_- dz_1^\ddagger, \qquad
a^\dagger = \int dz_1 {\mathbb  A}_+ , \ee i.e., ${\mathbb A}_- $
and ${\mathbb A}_+ $ fulfill the role of a linear density of the
annihilation $a$ and the creation $a^\dagger, $  respectively. We
notice that \be [ a , a^\dagger ]= \int \{ {\mathbb A}_- ,
{\mathbb A}_+ \} dz_1^\ddagger dz_1, \qquad \{ a , a^\dagger \}=
\int [ {\mathbb A}_- ,  {\mathbb A}_+ ] dz_1^\ddagger dz_1, \ee
i.e., the commutator and  anticommutator of the even operators $a$
and $a^\dagger$ are obtained by integrating, on  the entire odd
Grassmann space, the anticommutator and  commutator of the odd
operators ${\mathbb A}_- $ and $ {\mathbb A}_+ , $ respectively.
This suggests the following definitions of the densiy of identity
${\mathbb I}$ and of an energy type density  ${\mathbb H}: $  \be
{\mathbb I} = \{ {\mathbb A}_- , {\mathbb A}_+ \} = z_1
z_1^\ddagger, \qquad {\mathbb  H} = [ {\mathbb A}_- , {\mathbb
A}_+ ] = {w \over 2} z_1 z_1^\ddagger \{a , a^\dagger \}. \ee

As we know,  the eigenstates of the annihilation operator
correponding  to the complex eigenvalue $\alpha $ are given by the
standard harmonic oscillator coherent states $ |\alpha \rangle =
D(\alpha ) | 0 \rangle . $ They verify the eigenvalue equation \be
a |\alpha \rangle = \alpha |\alpha \rangle. \ee Multipliying both
sides of this equation by $z_1^\ddagger ,$ then integrating with
respect to this Grassmann variable and finally using
\eqref{den-anni-crea}, we get \be \int  {\mathbb  A}_- | \alpha
\rangle dz_1^\ddagger = \int   \alpha z_1^\ddagger | \alpha
\rangle dz_1^\ddagger, \ee i.e., by comparing both sides of this
last equation  we conclude that a class of eigenstates of the odd
operator ${\mathbb  A}_- $ corresponding to the $\alpha
z_1^\ddagger $ eigenvalue are given by the standard harmonic
oscillator coherent states $ \epsilon_\phi |\alpha \rangle.$

\subsection{Generalized  supersqueezed states}
Let us now solve the eigenvalue equation \eqref{eq-reduite}. A
class of solutions can be constructed firstly, by expressing
$|\psi \rangle $ in terms of a generalized $su(1,1)$ squeeze
operator (the normaliser of the H--W algebra), following this way
the construction of the standard squeezed states associated to the
simple harmonic oscillator system\cite{OrSa}. Indeed, let us write
\be \label{fononde-comprimee}|\psi \rangle = S( {\cal X}_0 )
|\varphi \rangle, \ee where the squeeze operator $S ({\cal X }_0 )
$ is given by \be S( {\cal X}_0 ) = \exp\left( {\cal X }_0 {
{(a^\dagger )}^2 \over 2 } - { {\cal X }_0 }^\ddagger  { a^2 \over
2 }\right),  \label{squeezed-su11}  \ee with ${\cal X}_0 $  an
even invertible Grassmann number, ${\cal X }_0^\ddagger $ its
adjoint (see Appendix \ref{sec-Not}).

Inserting \eqref{fononde-comprimee} in \eqref{eq-reduite}, using
the relation \be S^\ddagger ( {\cal X}_0 ) a S( {\cal X}_0 ) =
\cosh (\| {\cal X }_0 \|) \, a  + \sqrt{{\cal X}_0 }
{\left(\sqrt{{\cal X}_0^\ddagger } \right)}^{-1} \sinh (\| {\cal X
}_0 \|) \, a^\dagger,
 \ee
where $\| {\cal X }_0 \| = \sqrt{{\cal X }_0 {\cal X}_0^\ddagger},
$ and choosing ${\cal X }_0 $ in such a way that it satisfies \be
\sqrt{{\cal X}_0 } {\left(\sqrt{{\cal
X}_0^\ddagger}\right)}^{-1}\sinh (\| {\cal X }_0 \|) + \beta_0
\cosh (\| {\cal X }_0 \|) =0,  \label{gxbeta0} \ee we get the
following eigenvalue equation for $|\varphi \rangle :$  \beqa
\label{presque-impair} \biggl[ {\cal G} ({\cal X}_0 , \beta )  \,
a + \beta_1 \cosh (\| {\cal X }_0 \|) \, a^\dagger \biggr]
|\varphi \rangle = z |\varphi \rangle, \eeqa where \be {\cal G}
({\cal X}_0 , \beta ) = \cosh (\| {\cal X }_0 \|) + \beta \sqrt{
{\cal X}_0^\ddagger } {\left(\sqrt{\cal X}_0 \right)}^{-1} \sinh
(\| {\cal X }_0 \|). \ee Let us notice that this last coefficient
can be written on the form \be {\cal G} ({\cal X}_0 , \beta ) =
{\cal G} ({\cal X}_0 , \beta_0 )
 \left( \epsilon_\phi + \beta_1 \, {\left({\cal G}
({\cal X}_0 , \beta_0 ) \right) }^{-1} \sqrt{ {\cal X}_0^\ddagger
} {\left( \sqrt{\cal X}_0 \right)}^{-1} \sinh (\| {\cal X }_0 \|)
\right)  \label{gx0beta} \ee where, taking into account
\eqref{gxbeta0}, \be {\cal G} ({\cal X}_0 , \beta_0 ) = \left[
\epsilon_\phi - \beta_0^2 \, {\cal X}_0^\ddagger  {(  {\cal X}_0
)}^{-1} \right] \cosh (\| {\cal X }_0 \|).  \ee

Multiplying both sides of the equation \eqref{presque-impair} by
the inverse of ${\cal G} ({\cal X}_0 , \beta ) $ and taking into
account \eqref{gx0beta}, we get \be \label{vrai-reduction} [ a +
{\hat \beta}_1 a^\dagger ]|\varphi \rangle = {\hat z} |\varphi
\rangle, \ee where \be {\hat \beta}_1 = \beta_1 {\left( {\cal G}
({\cal X}_0 , \beta_0 ) \right)}^{-1}  \cosh (\| {\cal X }_0 \|)
\quad \in {\mathbb C}B_{L_{1}}, \ee and \be {\hat z } = \left[
{\left( {\cal G} ({\cal X}_0 , \beta_0 ) \right)}^{-1}  -  \beta_1
\sqrt{{( {\cal X}_0 }^\ddagger ) } {\left( \sqrt{\cal X}_0
\right)}^{-1} \sinh (\| {\cal X }_0 \|) \right]   z . \ee The
equation \eqref{vrai-reduction}  is thus  simpler to solve than
\eqref{eq-reduite}. Indeed, we can again try a solution of the
type \be \label{sol-gen} |\varphi \rangle = \sum_{n =0}^\infty C_n
| n \rangle, \qquad C_n \in {\mathbb C} B_L.  \ee Inserting it in
\eqref{vrai-reduction}, using the raising and lowering properties
of the operators $a^\dagger$ and $a, $ and the orthogonality
conditions of the states $\left\{ |n\rangle \right\},$  we get the
recurrence relation \be
 C_{n+1} = {  [{\hat z} C_n - \sqrt{n} {\hat
\beta}_1 C_{n-1}] \over \sqrt{n+1}}, \qquad n=2, \ldots, \ee with
\be  C_1 = {\hat z} C_0, \ee and $C_0 $ is an arbitrary constant.
Proceeding by iteration we get \be \label{recu-cenes} C_n = {1
\over \sqrt{n!}} \left( {\hat z}^n - \sum_{k=0}^{n-2} (k+1){\hat
z}^{(n-2-k)} {\left( {\hat z}^\ast \right)}^k \, {\hat \beta}_1
\right) \, C_0, \qquad n=2,3,\ldots. \ee This expression may be
written in a closed form. Indeed, as we can show that \beqa
\sum_{k=0}^{n-2} (k+1){\hat z}^{(n-2-k)} {\left( {\hat z}^\ast
\right)}^k &=& {n (n-1) \over 2! } {({\hat z}_0 )}^{n-2} - {n
(n-1)(n-2) \over 3! } {({\hat z}_0 )}^{n-3} {\hat z}_1  \\ &=& {1
\over 2!} {\partial^2 \over
\partial {\hat z}_0^2 } {({\hat z}_0 )}^n - {1 \over 3!}
{\partial^3 \over
\partial {\hat z}_0^3 } {({\hat z}_0 )}^n \, {\hat z}_1, \eeqa
 the  relation \eqref{recu-cenes} becomes \be C_n =
{1 \over \sqrt{n!}} \left( {\hat z}^n - \left[ {1 \over 2!}
{\partial^2 \over \partial {\hat z}_0^2 } {({\hat z}_0 )}^n - {1
\over 3!} {\partial^3 \over \partial {\hat z}_0^3 } {({\hat z}_0
)}^n \, {\hat z}_1 \right] {\hat \beta}_1 \right) \, C_0, \qquad
n=2,3, \ldots, \ee which is also valid  for $n=1.$ Finally,
inserting this result into \eqref{sol-gen}, and after some
manipulations we obtain a general solution of
\eqref{vrai-reduction}, which is \be
\label{sol-quasifinal}|\varphi \rangle= \left[ e^{{\hat z}_1
a^\dagger } - {\hat \beta}_1 { {( a^\dagger )}^2 \over 2! } +
{\hat z}_1 {\hat \beta}_1 { {( a^\dagger )}^3 \over 3! } \right]
e^{{\hat z}_0  a^\dagger } | 0 \rangle C_0 . \ee A normalized
version of \eqref{sol-quasifinal} is given by \be
\label{solvarphi} |\varphi \rangle = \exp\left[- {\hat \beta}_1 {
{( a^\dagger )}^2 \over 2 } - {\hat z}_1 {\hat \beta}_1 { {(
a^\dagger )}^3 \over 3 } \right] {\mathbb D} ({\hat z}_0 )
{\mathbb D} ({\hat z}_1 ) | 0 \rangle \, {\hat C} ({\hat z} ,
{\hat \beta}_1 ) ,  \ee where the operator ${\mathbb D}$ has been
defined in \eqref{mathd-super-adj}. The normalization constant
${\hat C }$ is given by \be {\hat C } ({\hat z} , {\hat \beta}_1 )
= {\left(\sqrt{\Gamma }\right)}^{-1} \left[ \epsilon_\phi + {1
\over 2} {\left(\sqrt{\Gamma }\right)}^{-1} \Omega
{\left(\sqrt{\Gamma }\right)}^{-1} \right], \ee with \be \Gamma
({\hat z} , {\hat \beta}_1 ) = \epsilon_\phi - {1 \over 2} \left(
{({\hat z}^\ddagger )}^2 {\hat \beta}_1 + {({\hat \beta}_1
)}^\ddagger {{\hat z}}^2 \right) - {1 \over 3} \left( {({\hat
z}^\ddagger )}^3 {\hat z}_1 {\hat \beta }_1 + {({\hat \beta}_1
)}^\ddagger  {({\hat z}_1)}^\ddagger {{\hat z}}^3 \right) \ee and
\beqa \nonumber \Omega ({\hat z} , {\hat \beta}_1 ) &=& \Biggl[ {1
\over 6} \left( {({\hat z}^\ddagger )}^3 {\hat z}_1  {{\hat z}}^2
+ {({\hat z}^\ddagger )}^2 {({\hat z}_1 )}^\ddagger {{\hat z}}^3
\right)
\\ &+& \nonumber
\left( {({\hat z}^\ddagger )}^2 {\hat z}_1  {{\hat z}} + {({\hat
z}^\ddagger )} {\hat z}_1 + {({\hat z}^\ddagger )} {({\hat z}_1
)}^\ddagger {{\hat z}}^2 + {({\hat z}_1 )}^\ddagger {{\hat z}}
\right) - {1 \over 4} \left( { ({\hat z}^\ddagger )}^2 {\hat z}^2
+ 4  {\hat z}^\ddagger  {{\hat z}} + 2 \right) \Biggr] {({\hat
\beta}_1 )}^\ddagger {\hat \beta }_1  \\ \nonumber &-& {1 \over 9}
\left( {({\hat z}^\ddagger )}^3 {\hat z}^3 + 9 {({\hat z}^\ddagger
)}^2 {\hat z}^2 + 24 {\hat z}^\ddagger  {\hat z} + 6 \right )
{({\hat z}_1 )}^\ddagger {\hat z}_1 {({\hat \beta}_1 )}^\ddagger
{\hat \beta }_1  \\ \nonumber &-& \left({({{\hat z}_0}^\ddagger
)}^2 {\hat \beta }_1 + {({\hat \beta}_1 )}^\ddagger {( {\hat z}_0
)}^2 \right) {({\hat z}_1 )}^\ddagger {\hat z}_1. \eeqa

From \eqref{fononde-comprimee} and \eqref{solvarphi} we conclude
that a class of normalized solutions of the eigenvalue equation
\eqref{eq-reduite}, corresponding to the eigenvalue $z,$ is given
by the generalized supersqueezed states \be
\label{gen-super-squeezed} |\psi \rangle = S( {\cal X}_0 )
\exp\left[- {\hat \beta}_1 { {( a^\dagger )}^2 \over 2 } - {\hat
z}_1 {\hat \beta}_1 { {( a^\dagger )}^3 \over 3 } \right] {\mathbb
D} ({\hat z}_0 ) {\mathbb D} ({\hat z}_1 ) | 0 \rangle \, {\hat C}
({\hat z} , {\hat \beta}_1 ) . \ee

Let us now give some examples of such states.

\subsubsection{Standard supersqueezed states} The standard
supersqueezed states are obtained from \eqref{gen-super-squeezed}
when $\beta_1 = 0$ and $z_1 = 0,$ i.e., when ${\hat \beta}_1 =0, $
${\hat z}_1 = 0 $ and ${\hat z}_0 =  {\left({\cal G} ({\cal X}_0 ,
\beta_0 ) \right) }^{-1} z_0 .$ They are given by \be |\psi
\rangle = S( {\cal X}_0 ) {\mathbb D} ({\hat z}_0 )  | 0 \rangle,
\ee where $ {\cal X}_0 $ and ${\hat z}_0 $ remain even Grassmann
valued numbers.

\subsubsection{A new class of supersqueezed states}
Another class of supersqueezed states  appears in
\eqref{gen-super-squeezed}, because of  the possibility  to choose
in \eqref{eq-reduite} a non zero odd component of the variable
$\beta.$  For example, if we choose $\beta_0 =0,$ i.e., ${\cal
X}_0 =0, \, {\hat \beta}_1 = \beta_1 $ and $ {\hat z} = z, $ then
from \eqref{gen-super-squeezed} we obtain the following class of
states \be |\psi \rangle = \exp\left[- \beta_1 { {( a^\dagger )}^2
\over 2 } -  z_1 \beta_1 { {( a^\dagger )}^3 \over 3 } \right]
{\mathbb D} ( z_0 ) {\mathbb D} ( z_1 ) | 0 \rangle {\hat C} ( z ,
\beta_1 ) . \ee They are obtained by applying  the operator \be
\exp\left[- \beta_1 { {( a^\dagger )}^2 \over 2 } - z_1 \beta_1  {
{( a^\dagger )}^3 \over 3 } \right]
 \ee to the generalized coherent states
 \eqref{gen-cohe-states} of $a .$ In the special case where $z_1 = 0,$
 we get to the normalized supersqueezed states
 \beqa \nonumber
 |\psi \rangle &=& \left[ \epsilon_\phi + {1 \over 4}  \beta_1^\ddagger
 \beta_1 \left( z_0^2 {( z_0^\ddagger )}^2 + 4 z_0 z_0^\ddagger
 +2 \right) \right] \exp\left[ - {1 \over 8}  \beta_1^\ddagger
 \beta_1 \left( a^2 {( a^\dagger )}^2 + {( a^\dagger )}^2 a^2
 \right) \right] \\
 && \exp\left[- \left( \beta_1 {{( a^\dagger )}^2 \over 2} -
 \beta_1^\ddagger {a^2 \over 2} \right) \right]
  {\mathbb D} ( z_0 ) | 0 \rangle,  \eeqa  which are written in terms of the
superunitary operator $S(- \beta_1) $  as defined in
\eqref{squeezed-su11}. Moreover, in the case where $\beta_1 \in
{\mathbb R}B_{L_{1}}, $ this last equation becomes \be |\psi
\rangle =
 S( -\beta_1 )
  {\mathbb D} ( z_0 ) | 0 \rangle,  \ee i.e.,
  we are in the presence of a class of supersqueezed  states which are constructed by applying
the superunitary  supersqueeze operator $S(- \beta_1 )$  to the
standard harmonic oscillator coherent states.

\section{SAES  associated to the fermionic superalgebra}
\label{sec-fersuper} In this section, we will construct the SAES
associated with the fermionic superalgebra generated by $\{b ,
b^\dagger, I\}$ which satisfy the non-zero supercommutation
relation \eqref{supercom-bbd} . The general eigenvalue equation
writes as \be [B_-  b + B_+ b^\dagger + B_3 I ] | \psi \rangle = Z
| \psi \rangle, \qquad B_\pm, \, Z \in {\mathbb C}B_L .
\label{fermionica} \ee Here we will distinguish again two cases:
firstly when $ B_+ = B_3 = 0 $ and secondly when $B_- $ is
invertible so that the equation \eqref{fermionica} reduces to \be
( b + \delta b^\dagger) |\psi \rangle = z |\psi \rangle, \qquad
\delta, z \ \in {\mathbb C}B_L . \label{bbd-squeezed-eq} \ee

\subsection{The b-fermionic eigenstates}
 Let us solve
 \be
\label{super-anh-e} B \; b |\psi \rangle = Z |\psi \rangle, \qquad
B, Z \ \in {\mathbb C}B_L . \ee  Since the fermionic graded Fock
space is reduced to the vectors $|- \rangle $ (even) and $ | +
\rangle $ (odd) which act as in \eqref{raising-b}, a solution of
\eqref{super-anh-e} writes as \be \label{super-sol} |\psi \rangle
=  C | - \rangle + D |+ \rangle, \qquad C , D \ \in {\mathbb
C}B_L. \ee Inserting (\ref{super-sol}) into (\ref{super-anh-e})
and using (\ref{raising-b}), we get \be B D^\ast | - \rangle = Z C
|- \rangle + Z  D |+ \rangle. \ee The orthogonality  of the states
$|- \rangle $ and $ | + \rangle $ leads to the following set of
algebraic equations \beqa \nonumber
B D^\ast &=& Z  C \\
 Z  D & =& 0,  \label{sistema1}
 \eeqa
or by conjugation of the first one, \beqa \nonumber
B^\ast D &= & Z^\ast C^\ast  \\
Z D &=&0. \label{sistem-2} \eeqa Let us mention that, when $B_\phi
\ne 0, $ we have evidently the normalized  solution $ |\psi\rangle
= | - \rangle $ when the eigenvalue $Z$ is zero, but due to the
presence of Grassmann value quantities, when $B_\phi = 0, $ we
have a larger set of solutions. For instance, for $B=B_1,$ we
find,  a solution of the form \be \label{simple-supercohe} |\psi
\rangle = C | - \rangle \pm B_1 | + \rangle. \ee Normalized
eigenstates are given by \be | \psi \rangle = \exp\left[ \pm
\left( B_1 b^\dagger + B_1^\ddagger b \right)\right] |- \rangle.
\ee

When  $Z \ne 0,$ non-trivial solutions appears if and only if $
Z_\phi =0. $ From \eqref{sistem-2}, we have $ D_\phi = 0.$ To
solve completely  the system \eqref{sistem-2} we have to
distinguish two cases.

If $B_\phi \ne 0, $ we can solve $D$ from the first equation of
(\ref{sistem-2}) \be \label{disole} D = {( B^\ast )}^{-1} Z^\ast
C^\ast = {\left( B^{-1} Z \right)}^\ast C^\ast =z^\ast C^\ast, \ee
where $ z =z_0 + z_1 = \left( B^{-1} Z \right).$ Now inserting
(\ref{disole}) into the second equation of
 \eqref{sistem-2}, we get \be Z z^\ast C^\ast = 0. \ee
 Normalized solutions will be obtained if $C_\phi \ne 0 $ and we thus get
  \be Z z^\ast = 0, \ee which can be written explicitly \be \label{condiciones} z_0^2 =0, \qquad z_0 Z_1 =
z_1 Z_0. \ee

The normalized eigenstates of $ Bb $ with the eigenvalue $Z$
satisfying \eqref{condiciones} are given by \be |\psi \rangle =
\biggl( | - \rangle +  z^\ast    | + \rangle \biggr) C, \ee where
$C$ is an arbitrary Grassmann number such that $C_\phi \ne 0 .$
They can be written as \be |z_0 ; z_1 \rangle = {\mathbb T} (z_1)
{\mathbb T} (z_0)  | - \rangle, \label{b-saes}  \ee where the
superunitary operators $\mathbb T $ are given by \be
 {\mathbb T} ( z_1 ) = \exp\left( b^\dagger z_1 -  z_1^\ddagger b \right), \qquad
 {\mathbb T} ( z_0 ) = \exp\left( z_0 b^\dagger -
z_0^\ddagger b \right). \ee The $b$--SAES  are obtained from
\eqref{b-saes} when $ B = \epsilon_\phi, $ so that $z_0 = Z_0$ and
$z_1 =Z_1. $ We notice that when $z_0 = 0, $ they reduces to the
standard supercoherent states associated to the system
characterized by the fermionic Hamiltonian $ H = b^\dagger b - {1
\over 2 }.$

If $B_\phi =0,$ the problem is a little more tricky. We can write
(\ref{sistema1}) explicitly as \beqa
B_0 d_0 - B_1 d_1 &=& Z_0 c_0 + Z_1 c_1 \label{sistema-a} \\
B_1 d_0 - B_0 d_1 &=& Z_1 c_0 + Z_0 c_1 \label{sistema-b} \\
Z_0 d_0 + Z_1 d_1 &=& 0 \label{sistema-c} \\
Z_1 d_0 + Z_0 d_1 &=& 0, \label{sistema-d} \eeqa where we have
taken $C = c_0 + c_1 $ and $ D= d_0 + d_1. $ In this way, for
instance, when $B_0 \ne 0 $ and $ {(B_0)}^2 \ne 0, $ we can
combine \eqref{sistema-a} and \eqref{sistema-b} to obtain \beqa
{(B_0)}^2 d_0 &=& (B_0 Z_0 - B_1
Z_1 ) c_0 + (B_0 Z_1 - B_1 Z_0 ) c_1 , \label{sistema-ab1} \\
{(B_0)}^2 d_1 &=& (B_1 Z_0 - B_0 Z_1 ) c_0 + (B_1 Z_1 - B_0 Z_0 )
c_1 \label{sistema-ab2}\eeqa and then combine this last system of
equations with \eqref{sistema-c} and \eqref{sistema-d} to get
\beqa
 Z_0 \left( 2 B_1   Z_1 - B_0  Z_0 \right)   c_0  + B_1 {(Z_0)}^2
c_1 =0, \label{sistema-ab12-c} \\
Z_0 \left( 2 B_1   Z_1 - B_0  Z_0 \right)  c_1  +  B_1 {(Z_0)}^2
c_0 =0.   \label{sistema-ab12-d} \eeqa The systems
(\ref{sistema-ab1}-\ref{sistema-ab2}) and
(\ref{sistema-ab12-c}-\ref{sistema-ab12-d}) are equivalent to \be
{(B_0)}^2 D =  B  Z^\ast C^\ast \label{D-C} \ee and \be Z_0 \left(
2 B_1 Z_1 - B_0  Z_0 +  B_1 Z_0  \right) C = 0, \ee respectively.
As we search for normalized solutions, we must take $C_\phi \ne 0.
$ This implies the following condition for the $Z$ eigenvalue:
\beqa \label{Zcon-a}
Z_0 \left( 2 B_1   Z_1 -   B_0  Z_0  \right) &=& 0 \\
B_1 {(Z_0)}^2 &=&0.  \label{Zcon-b} \eeqa  Then, the normalized
eigenstates of (\ref{super-anh-e}) corresponding to the $Z$
eigenvalue  satisfying (\ref{Zcon-a}-\ref{Zcon-b}) are given by
(\ref{super-sol}), with $C$ an arbitrary Grassmann number such
that $ C_\phi \ne 0, $ and $D$  verifying  \eqref{D-C}.

Following a similar procedure,  when $B_0 = 0$ and $B_1 \ne 0, $
the normalized  solutions of (\ref{super-anh-e}) corresponding to
the $Z$ eigenvalue satisfying the conditions \be {(Z_0)}^2 =0,
\qquad Z_0 Z_1 = 0, \ee are given by  \eqref{super-sol}, with
$C_\phi \ne 0, $ and $D$  verifying \be B_1 D = - Z^\ast C^\ast .
\ee When $B_0 \ne 0$ et $B_1=0, $ the  solutions corresponding to
the $Z$ eigenvalue satisfying the conditions \be {(Z_0)}^2 =0, \ee
are given by  \eqref{super-sol}, with $C_\phi \ne 0, $ and $D$
verifying \be B_0 D = Z^\ast C^\ast. \ee

Other classes of solutions can be reached  by imposing other
conditions on the  coefficient $B.$

\subsection{Supersqueezed states}
 Let us now solve  the eigenvalue  \eqref{bbd-squeezed-eq}. If we assume again a solution of the type
 \eqref{super-sol}, then by inserting it in \eqref{bbd-squeezed-eq}, using the
raising and lowering properties \eqref{raising-b} and  the
orthogonality between the sates $|-\rangle$ and $|+\rangle,$ we
get the following algebraic  Grassmann equations for determining
$C$ and $D :$
\beqa \label{dconju} D^\ast &=& z C, \\
\label{betaconju}  \delta C^\ast &=& z D. \eeqa By conjugating the
equation \eqref{dconju} and then by inserting it in
\eqref{betaconju}, we get  \be \label{zetabeta} ( z z^\ast -
\delta ) C^\ast =0. \ee As we are interested  in normalized
solutions, we must take $C_\phi \ne 0,  $ then  \eqref{zetabeta}
implies: \be \label{condi-zeta0} z_0^2 = \delta, \ee   that is,
$\delta $ is an even Grassmann number. Inserting \eqref{dconju} in
\eqref{super-sol}  and considering the conditions
\eqref{condi-zeta0}, we conclude that a set of normalized
eigentates of the operator $ ( b + \delta_0 b^\dagger )$
corresponding  to the eigenvaleue $ z = \pm \sqrt{\delta_0 }  +
z_1 $ is given by \be |\delta_0 , z_1 \rangle^\pm  = \biggl(| -
\rangle -  ( z_1 \mp  \sqrt{\delta_0} ) |+ \rangle \biggr) C. \ee
It is not too hard to show that the corresponding normalized
supersqueezed states are given by \be \label{super-fer-comprimes}
|\delta_0 , z_1 \rangle^\pm = \exp\left( b^\dagger  z_1 -
z_1^\ddagger b \right) \exp\left[\pm \sqrt{\delta_0 } \left(
b^\dagger + z_1^\ddagger \right) \right] | - \rangle \, N^\pm
(\delta_0 , z_1), \ee where the normalization constant $N^\pm $ is
given by \be N^\pm (\delta_0 , z_1) = {\cal F}^{-1} \ \left[
\epsilon_\phi \mp  {1 \over 2 } {\cal F}^{-1} \left(
\sqrt{\delta_0} z_1^\ddagger + {(\sqrt{\delta_0})}^\ddagger z_1
\mp \sqrt{\delta_0} {(\sqrt{\delta_0})}^\ddagger z_1^\ddagger z_1
\right) {\cal F}^{-1} \right], \ee with \be {\cal F} (\delta_0) =
\sqrt{ 1 +\sqrt{\delta_0} {(\sqrt{\delta_0})}^\ddagger }. \ee We
notice that in the limit $\delta_0 \mapsto 0$ the supersqueezed
states \eqref{super-fer-comprimes} becomes the eigenstates of the
operator $b$ corresponding to the eigenvalue $z=z_1 .$

\section{SAES  associated to the  Heisenberg--Weyl Lie superalgebra}
\label{sec-super-hw-lie-super}
 Let us now compute the SAES associated
to the H-W Lie superalgebra generated by the set of generators
$\{a, a^\dagger, I , b ,b^\dagger \}$ whose non zero
super-commutation relations are given by the relations
\eqref{com-aad} and \eqref{supercom-bbd}. The eigenvalue equation
is written as \be \label{gen-saes-eq} [ A_- a + A_+ a^\dagger +
A_3 I + B_- b + B_+ b^\dagger ] |\psi \rangle = Z |\psi \rangle,
\qquad A_\pm, A_3, B_\pm, Z \in {\mathbb C}B_L. \ee Here we
concentrate in the case where $ {(A_- )}_\phi \ne 0, $ i.e., $A_-
$ is an invertible Grassmann number. In this case, we can express
\eqref{gen-saes-eq} in the form \be \label{gen-super-eq} [ a +
\beta  a^\dagger + \gamma b + \delta b^\dagger ] |\psi \rangle = z
|\psi \rangle, \qquad \beta, \gamma, \delta, z \, \in {\mathbb
C}B_L. \ee

Special cases of this problem have been  considered in sections
\ref{sec-hwla} and \ref{sec-fersuper}. Here we consider the cases
where we have the presence of both bosonic and fermionic operators
in the eigenvalue equation \eqref{gen-super-eq}.

\subsection{Generalized supercoherent states}
First, we take the particular eigenvalue equation  \be
\label{super-cohe-equa} [ a + \gamma b ] |\psi \rangle = z |\psi
\rangle, \qquad \gamma, z \in {\mathbb C}B_L. \ee Let us assume a
solution of the type \be \label{sol-type-bos-fer}|\psi \rangle =
\sum_{n=0}^\infty \biggl( C_n |n ; - \rangle + D_n | n ; + \rangle
\biggr), \ee where $C_n, \, D_n \in {\mathbb C}B_L .$ By inserting
(\ref{sol-type-bos-fer}) in (\ref{super-cohe-equa}), using the
lowering properties of operators $a$ and $b,$ Eqs.
\eqref{raising-a} and \eqref{raising-b}, and the orthogonality
properties  of the  graded Fock space basis $ \{  |n ; - \rangle ,
|n ; + \rangle, n \in {\mathbb N}  \}, $ we get the recurrence
relations  \beqa \label{recu-aa} \sqrt{n+1} C_{n+1} + \gamma D^{\ast}_n = z C_n, \\
\label{recu-bb}\sqrt{n+1}  D_{n+1} = z D_n. \eeqa From
\eqref{recu-bb}, it is  easy to find  the expression of the
coefficients $D_n  $ in terms of an arbitrary constant  $D_0 :$
\be \label{dnd0} D_n = {z^n \over \sqrt{n!}} D_0, \quad n=1,2,
\ldots. \ee   Then, by inserting (\ref{dnd0}) in (\ref{recu-aa}),
we get  the following recurrence relation for the coefficients
$C_n :$ \be C_{n+1} = {1\over \sqrt{n+1}} \left[ z C_n -   \gamma
 {{(z^\ast)}^n \over \sqrt{n!}} D^\ast_0 \right], \qquad n=0,1,2,\ldots.  \ee
Finally, proceeding by iteration we get \be C_n= {1 \over
\sqrt{n!}} \left[ z^n C_0 - \left( \sum_{k=0}^{n-1} z^{(n-1-k)}
\gamma {(z^\ast)}^k \right) D^\ast_0 \right], \qquad n=1,2,\ldots,
\ee where $C_0 $ is an arbitrary constant. Since $C_0 $ and $D_0 $
are arbitrary constants, the equation (\ref{sol-type-bos-fer})
gives two independent solutions. The first one consists of  the
standard coherent states \be \label{cohe-n;0}| z ; - \rangle =
\sum_{n=0}^\infty {z^n \over \sqrt{n!}} C_0 | n ; - \rangle . \ee
To find the second one, we use the formula \be {1\over n+1 }
\sum_{k=0}^{n} z^{(n-k)} \gamma {(z^\ast)}^k = ( \gamma_0 {z_0 }^n
+ z^n \gamma_1). \ee We thus get the generalized coherent states
on the form \beqa  \widetilde{| z , \gamma ; + \rangle} =
\widetilde{| z, \gamma_0 , \gamma_1 ; + \rangle } &=& \left[
\sum_{n=0}^\infty { z^n \over \sqrt{n!}} | n ; + \rangle -
a^\dagger \sum_{n=0}^\infty {\left( \gamma_0 {z_0 }^n + z^n
\gamma_1 \right ) \over \sqrt{n!}} | n ; - \rangle \right]
D^\ast_0 \nonumber \\ &=& \exp\left[- \left(\gamma_0 (1+ z_1
a^\dagger) + \gamma_1 \right) a^\dagger b  \right] e^{z a^\dagger}
|0 ; + \rangle \ D^\ast_0 \label{etat-tilde}.\eeqa The normalized
version of the states (\ref{cohe-n;0}) is  given by \be | z  ; -
\rangle = | z_0 , z_1 ; - \rangle = {\mathbb D} (z_0) {\mathbb D}
(z_1)  | 0 ; - \rangle. \label{set-coh-states}\ee It is similar to
the one obtained in \eqref{gen-cohe-states}. A set of normalized
generalized supercoherent states, orthogonal to
\eqref{set-coh-states} is given by the formula \be | z , \gamma ,
+ \rangle = | z_0 , z_1 , \gamma_0 , \gamma_1 ; + \rangle = {
\widetilde{| z , \gamma_0 , \gamma_1 , + \rangle} - | z ; -
\rangle \ \langle - ; z \widetilde{| z , \gamma_0 , \gamma_1 , +
}\rangle \over || \, \widetilde{| z , \gamma_0 , \gamma_1 , +
\rangle} - | z ; - \rangle \ \langle - ; z \widetilde{| z ,
\gamma_0 , \gamma_1 , + \rangle } \, ||}. \ee  After some
calculations,  we get the set of generalized supercoherent states
\beqa \nonumber | z_0  , z_1 , \gamma_0 , \gamma_1 ; + \rangle &
=& {\mathbb D } ( z_0 ) {\mathbb D} (z_1 )
 \Biggl\{  | 0 ; + \rangle \\ \nonumber &-& \biggl[ \left( 1 - {
1 \over 2} z_1^\ddagger z_1 \right) {\mathbb D} (- z_1 ) (
a^\dagger + z_0^\ddagger ) \gamma_0 e^{z_1 z_0^\ddagger } + ( 1 +
z_1^\ddagger z_1 ) a^\dagger \gamma_1   \\  &-&   ( 1 -
z_1^\ddagger z_1 )
 z^\ddagger \gamma_0 e^{z_1 z_0^\ddagger }   \biggr]|0 ; - \rangle
\Biggr\} N (z_0 , z_1 , \gamma_0 , \gamma_1 ),
\label{gen-super-cohe} \eeqa where the normalization  constant $ N
$ is given by \be \label{nor-factor}
 N (z_0 , z_1 , \gamma_0 , \gamma_1 ) = {\cal B}^{-1}
 \left[ 1 - {\cal B }^{-1} \left( \gamma_1^\ddagger \gamma_1 -  \gamma_0^\ddagger \gamma_0
  {(z_0^\ddagger z_0 )}^2  \right)  z_1^\ddagger z_1 {\cal  B }^{-1} \right],
\ee with \be {\cal B} (\gamma_0 , \gamma_1 ) = \sqrt{1 +
\gamma^\ddagger \gamma } = \sqrt{1 + \gamma_0^\ddagger \gamma_0 +
\gamma_0^\ddagger \gamma_1 + \gamma_1^\ddagger \gamma_0 +
\gamma_1^\ddagger \gamma_1}. \label{cal-B} \ee

\subsubsection{Super coherent states}
The supercoherent states \eqref{gen-super-cohe} constitute  a
generalization of the super coherent states found by Aragone and
Zypman\cite{ArZy}. Indeed, from equations
(\ref{gen-super-cohe}-\ref{cal-B}) we see that, in the case where
$\gamma_1=0 $ and $ z_1 = 0, $ we have \be | z_0  , 0 , \gamma_0 ,
0 ; +  \rangle = {\left(\sqrt{1 + {\gamma_0}^\ddagger \gamma_0 }
\right) }^{-1} {\mathbb D} ( z_0 ) \ \biggl( |0 ; + \rangle -
\gamma_0 a^\dagger | 0 ; - \rangle \biggr) . \ee

\subsubsection{Other classes of supercoherent states}
Now if in  (\ref{gen-super-cohe}-\ref{cal-B}), we take $\gamma_0 =
0$ and $ z_0 = 0, $ we get \be | 0 , z_1 , 0 , \gamma_1 ; +
\rangle = \left( 1 - {1\over 2} \gamma_1^\ddagger \gamma_1 -
\gamma_1^\ddagger \gamma_1 z_1^\ddagger z_1 \right) {\mathbb D} (
z_1 ) \, \biggl( |0;+ \rangle - (1 + z_1^\ddagger z_1 ) a^\dagger
\gamma_1 | 0; - \biggr). \ee

We can also distinguish the case where $\gamma_1 = 0 $ and $z_0
=0.$ We get \beqa |  0 , z_1 , \gamma_0 , 0 ; + \rangle &=&
{\left(\sqrt{1 + {\gamma_0}^\ddagger \gamma_0 } \right) }^{-1}
{\mathbb D} ( z_1 ) \nonumber \\ & & \left\{ |0 ; + \rangle +
\gamma_0 \left[ \left( { z_1^\ddagger z_1 \over 2} - 1 \right)
 {\mathbb D} ( - z_1 ) a^\dagger + z_1^\ddagger  \right] | 0 ;
 - \rangle \right\}.  \eeqa

\subsubsection{Standard supercoherent states} In the case where
$\gamma = 0,$  \eqref{gen-super-cohe} becomes the standard
coherent states \be | z ; +  \rangle = | z_0 , z_1 ; + \rangle =
{\mathbb D} (z_0) {\mathbb D} (z_1)  | 0 ; + \rangle
\label{set-coh-states-1}. \ee By  combining  the two independent
solutions \eqref{set-coh-states} and \eqref{set-coh-states-1}, we
can construct a  solution of the type \be
 | z ; \rho, \tau  \rangle = \rho | z ; - \rangle + \tau | z ; +
 \rangle, \label{s-rotau}
\ee where  $\rho$ and $ \tau $ are  Grassmann numbers  such that
$\rho_1 z_1 = \tau_1 z_1 =0. $ Thus the states \eqref{s-rotau} are
eigenstates of $a$ corresponding to the eigenvalue $z.$ In
particular, if we take for example $ \rho = 1 - {z_1^\ddagger z_1
\over 2} $ and $ \tau = - z_1 ,$ then we obtain the supercoherent
states \be | z \rangle = {\mathbb D} (z_0) {\mathbb D} (z_1)
{\mathbb T} (z_1) | 0 ; - \rangle. \ee Moreover, if we take   $
z_1 =0,$ $ \rho = 1 - {\theta_1^\ddagger \theta_1 \over 2} $ and
$\tau = - \theta_1,$ we get the  standard supercoherent states
associated to the supersymmetric harmonic oscillator
\cite{YBeVh,Faetal} \be | z_0 , \theta_1 \rangle = {\mathbb D}
(z_0) {\mathbb T} (\theta_1 ) | 0 ; - \rangle.
\label{nieto-sup-coh} \ee

\subsection{Generalized supersqueezed states}
\label{sec-gen-gamma-delta}
Let us  now   find  the SAES
associated to the sub-superalgebra $\{a , b , b^\dagger, I \}. $
If the coefficient of $a$ in the linear combination is invertible
the problem reduces to solve the eigenvalue equation: \be
\label{super-a-b-bd} [a + \gamma b + \delta b^\dagger ] |\psi
\rangle = z |\psi \rangle, \qquad \gamma, \delta \in {\mathbb
C}B_L .\ee

We can show, see  Appendix \ref{sec-caso-general} section
\ref{sec-gen-super-squee-oper}, that  two classes of independent
solutions of the eigenvalue equation \eqref{super-a-b-bd} exist
and are given by \be \label{sup-sqee-c-a-b-bd} | \psi ; - \rangle
= \biggl[ \sum_{\ell \; {\rm even}}^\infty {\cal O}_{{a^\dagger}}
(\ell, \gamma, \delta^\ast, z_1 ) e^{z a^\dagger} | 0 ; - \rangle
- \sum_{\ell \; {\rm odd}}^\infty
  {\cal O}_{{a^\dagger}} (\ell, \delta, \gamma^\ast, z_1
) e^{za^\dagger} | 0; + \rangle \biggr] C_0 \ee and \be
\label{sup-sqee-d-a-b-bd}| \psi ; + \rangle = \biggl[ \sum_{\ell
\; {\rm even}}^\infty  {\cal O}_{{a^\dagger}} (\ell, \delta,
\gamma^\ast, z_1 ) e^{z a^\dagger} | 0 ; + \rangle - \sum_{\ell \;
{\rm odd}}^\infty   {\cal O}_{{a^\dagger}} (\ell, \gamma,
\delta^\ast, z_1 ) e^{za^\dagger} | 0; - \rangle \biggr] D_0^\ast,
\ee where $C_0$ and $D_0^\ast $ are arbitrary and invertible
Grassmann constants and \beqa \nonumber {\cal O}_{{a^\dagger}}
(\ell, \gamma, \delta^\ast, z_1) &=& {1\over \ell !} \Biggl\{
\stackrel{ \ell \; {\rm factors }}{\overbrace{(\gamma \delta^\ast
\gamma \delta^\ast \cdots )}} \left(  {(a^\dagger )}^\ell   - z_1
{(a^\dagger
)}^{\ell+1} \right) \\
 &+& { 1 \over \ell + 1 } \sum_{j=0}^{\ell} {(-1)}^{ j + \ell}
\stackrel{ (\ell - j ) \; {\rm factors }}{\overbrace{(\gamma
\delta^\ast \gamma \delta^\ast \cdots )}} z_1 \stackrel{  j \;
{\rm factors }}{\overbrace{(\cdots \gamma \delta^\ast \gamma
\cdots )}} {(a^\dagger )}^{\ell+1}
 \Biggr\} , \label{calo-oper-diff-rep} \eeqa
where $\ell = 0,1,2, \ldots. $

The superstates \eqref{sup-sqee-c-a-b-bd} and
\eqref{sup-sqee-d-a-b-bd} can be written in the form of  a
supersqueeze operator acting on the supercoherent state, that is
\beqa \nonumber | \psi ; - \rangle &=& {\cal O}_{ {\rm even}}
(a^\dagger ,\gamma, \delta^\ast, z_1) \exp\biggl[ - {\biggl( {\cal
O}_{ {\rm even}}  (a^\dagger , \gamma, \delta^\ast, z_1)
\biggr)}^{-1}  \\ && \biggl({\cal O}_{ {\rm odd}} (a^\dagger ,
\delta, \gamma^\ast, z_1)\biggr)
  e^{2 z_1 a^\dagger} b^\dagger \biggr] {\mathbb D} (z_0) {\mathbb
  D}(z_1) \ |0;- \rangle \ {\tilde C}_0 ,
  \eeqa
\beqa \nonumber | \psi ; + \rangle &=& {\cal O}_{ {\rm even}}
(a^\dagger ,\delta, \gamma^\ast, z_1)  \exp\biggl[ - {\biggl(
{\cal O}_{ {\rm even}}  (a^\dagger , \delta, \gamma^\ast, z_1)
\biggr)}^{-1}  \\ && \biggl({\cal O}_{ {\rm odd}} (a^\dagger ,
\gamma, \delta^\ast, z_1)\biggr)
  e^{2 z_1 a^\dagger} b \biggr] {\mathbb D} (z_0) {\mathbb
  D}(z_1) \ |0; + \rangle \ {\tilde D}_0^\ast ,
  \eeqa where \be \label{caleven}
 {\cal O}_{ {\rm even}} (a^\dagger , \gamma, \delta^\ast, z_1)  = \sum_{\ell \; {\rm even}}^\infty {\cal O}_{{a^\dagger}}
(\ell, \gamma, \delta^\ast, z_1 )\ee and \be \label{calodd}
 {\cal O}_{ {\rm odd}} (a^\dagger , \gamma, \delta^\ast, z_1)  = \sum_{\ell \; {\rm odd}}^\infty {\cal O}_{{a^\dagger}}
(\ell, \gamma, \delta^\ast, z_1 ).\ee

\subsubsection{Standard superqueezed states} In the case where
\label{sec-standar} $\gamma $ and $\delta$ are odd Grassmann
numbers, that is when $\gamma=\gamma_1$ and $\delta = \delta_1 ,$
it is easy to see from \eqref{calo-oper-diff-rep} that, the non
zero ${\cal O}_{{a^\dagger}} $ operators in
\eqref{sup-sqee-c-a-b-bd} and \eqref{sup-sqee-d-a-b-bd}
corresponds to \beqa \nonumber {\cal O}_{{a^\dagger}} (0, \gamma_1
, - \delta_1 , z_1 ) &=& 1, \quad {\cal O}_{{a^\dagger}} (1,
\delta_1 , - \gamma_1 , z_1 ) = \delta_1 a^\dagger - 2 \delta_1
z_1 {(a^\dagger)}^2, \\ \quad {\cal O}_{{a^\dagger}} (2, \gamma_1
, - \delta_1 , z_1 ) &=& - {1 \over 2!} \gamma_1 \delta_1
{(a^\dagger)}^2, \eeqa and \beqa \nonumber
 {\cal O}_{{a^\dagger}} (0, \delta_1 , - \gamma_1 , z_1 ) &=& 1, \quad
{\cal O}_{{a^\dagger}} (1, \gamma_1 , - \delta_1 , z_1 ) =
\gamma_1 a^\dagger - 2 \gamma_1 z_1 {(a^\dagger)}^2, \\
  {\cal O}_{{a^\dagger}} (2, \delta_1 , - \gamma_1 , z_1 ) &=& - {1 \over
2!} \delta_1 \gamma_1   {(a^\dagger)}^2, \eeqa respectively. By
inserting this results in \eqref{sup-sqee-c-a-b-bd} and
\eqref{sup-sqee-d-a-b-bd}, and after some simple manipulations, we
get the supersqueezed   states \be \label{odd-c-sol} | \psi ; -
\rangle = \exp\left[- {1\over 2} \gamma_1 \delta_1 {(a^\dagger
)}^2 \right] e^{ - \delta_1 a^\dagger b^\dagger } e^{z a^\dagger }
|0 ; - \rangle \ C_0 , \ee and \be \label{odd-d-sol}|\psi ; +
\rangle = \exp\left[- {1\over 2} \delta_1 \gamma_1 {(a^\dagger
)}^2 \right] e^{ - \gamma_1 a^\dagger b } e^{za^\dagger} |0 ; +
\rangle \ D_0^\ast ,\ee which are eigenstates of $ a + \gamma_1 b
+ \delta_1 b^\dagger. $ In these last expressions, we notice the
action of an normalizer operator acting on the corresponding
supercoherent states. The normalizer in equation \eqref{odd-c-sol}
transforms the algebra element $ a + \gamma_1 b + \delta_1
b^\dagger $ into $ a + \gamma_1 b $ whereas the normalizer in
equation \eqref{odd-d-sol} transforms it into $ a + \delta_1
b^\dagger. $ In fact, a complete reduction into the element $a$
only can be obtained. For instance, that is the case if we
multiply the normalizer in equation \eqref{odd-c-sol} by the
corresponding normalizer of the equation \eqref{etat-tilde} in the
special case where $\gamma_0=0, $   that is, by $e^{- \gamma_1
a^\dagger  b} .$ Moreover, if we consider the algebra element $ a
+ \beta_0 a^\dagger + \gamma_1 b + \delta_1 b^\dagger,$  a
normalizer operator transforming it into the element $a$ is given
by the standard supersqueeze operator\cite{BuRaRa} \be
\label{stand-super-sqee}{\bold G} (\beta_0, \gamma_1, \delta_1) =
\exp\left[ - \left( \beta_0 + \gamma_1 \delta_1 \right)
{{(a^\dagger )}^2 \over 2} \right] \exp\left( - \delta_1 a^\dagger
b^\dagger \right) \exp\left( - \gamma_1 a^\dagger b \right). \ee
In this way,  using the algebra eigenstates \eqref{nieto-sup-coh}
of the $a$ annihilator, we observe that a class of superalgebra
eigenstates of $ a + \beta_0 a^\dagger + \gamma_1 b + \delta_1
b^\dagger,$  corresponding to the eigenvalue $z_0, $ is given by
\be {\bold G} (\beta_0, \gamma_1, \delta_1) {\mathbb D}(z_0)
{\mathbb T}(\theta_1) |0 ; - \rangle C_0. \ee We notice that,
these supersqueezed states are obtained by acting with a
supersqueeze operator that is an element of the $OSP(2/2)$
supergroup on the supercoherent states associated to the
supersymmetric harmonic oscillator.  In this way, these SAES of
the algebra element $ a + \beta_0 a^\dagger + \gamma_1 b +
\delta_1 b^\dagger,$ are comparable to the supersqueezed states
for the supersymmetric harmonic oscillator \cite{KoNiTr,MMNieto}.

\subsubsection{\boldmath Spin ${1\over 2}$ representation AES  structure}
\label{sec-spin-unmedio} Let us consider now the special case
where both $\gamma$ and $\delta$ are even invertible Grassmann
numbers. Let us write $\gamma=\gamma_0  $ and $\delta = \delta_0.$
In this case, from \eqref{calo-oper-diff-rep}, we obtain \be
\label{calo-g0d0} {\cal O}_{{a^\dagger}} (\ell, \gamma_0 ,
\delta_0 , z_1) =\cases{ {{(a^\dagger)}^\ell \over \ell !}
{(\gamma_0 \delta_0)}^{\ell / 2} \exp\left( - {\ell \over \ell +1
} z_1 a^\dagger \right), &  if \, $\ell$ \, is \, even \cr
{{(a^\dagger)}^\ell \over \ell !} {(\gamma_0 \delta_0)}^{(\ell -1)
/ 2} \gamma_0 \exp\left( - z_1 a^\dagger \right), & if \, $\ell$
\, is \,  odd \cr}. \ee Thus, by inserting these results in
\eqref{caleven} and \eqref{calodd},  we get \beqa \nonumber
 {\cal O}_{ {\rm even}} (a^\dagger , \gamma_0, \delta_0, z_1)  &=& \sum_{\ell \; {\rm even}}^\infty
{{(  \sqrt{\gamma_0 \delta_0 } \, a^\dagger )}^\ell \over  \ell !}
\exp\left( - {\ell \over \ell +1 } z_1 a^\dagger \right) \\
\nonumber &=& \cosh(\sqrt{\gamma_0 \delta_0} \, a^\dagger) e^{-
z_1 a^\dagger } \\ & & \exp\left[ z_1 {(\sqrt{\gamma_0
\delta_0})}^{-1} {\left(\cosh(\sqrt{\gamma_0 \delta_0} \,
a^\dagger)\right)}^{-1} \sinh(\sqrt{\gamma_0 \delta_0} \,
a^\dagger)\right] \eeqa and \beqa \nonumber {\cal O}_{ {\rm odd}}
(a^\dagger , \gamma_0 , \delta_0 , z_1) &=&
{(\sqrt{\delta_0})}^{-1} \sqrt{\gamma_0} \sum_{\ell \; {\rm
odd}}^\infty {{(\sqrt{\gamma_0 \delta_0} \, a^\dagger )}^\ell
\over \ell !} \exp\left( - z_1 a^\dagger \right)
\\ &=&  {(\sqrt{\delta_0})}^{-1} \sqrt{\gamma_0} \, \sinh(\sqrt{\gamma_0
\delta_0} \, a^\dagger) \exp\left( - z_1 a^\dagger \right).
  \eeqa

By inserting these results in \eqref{sup-sqee-c-a-b-bd} and
\eqref{sup-sqee-d-a-b-bd} and   after some manipulations, we get
the set of independent eigenstates of $a + \gamma_0 b + \delta_0
b^\dagger :$ \beqa \nonumber |\psi ; - \rangle &=& \exp\left[-z_1
\left( a^\dagger - {(\sqrt{\gamma_0 \delta_0})}^{-1} T_h(\gamma_0,
\delta_0, a^\dagger)\right) \right]
 \cosh\Biggl\{ \sqrt{\gamma_0 \delta_0} \, a^\dagger - \\
 & &  {(\sqrt{\gamma_0})}^{-1}
\sqrt{\delta_0 }  \biggl[ 1   +     z_1 \biggl( 2a^\dagger -
 {(\sqrt{\gamma_0 \delta_0})}^{-1}  T_h(\gamma_0, \delta_0, a^\dagger)
 \biggl)
 \biggr] b^\dagger \Biggr\}  \,  e^{z a^\dagger}  |0; -\rangle \, C_0  \; \; \;
 \label{sup-sqee-c-z1} \eeqa
and \beqa \nonumber |\psi ; + \rangle &=& \exp\left[-z_1 \left(
a^\dagger - {(\sqrt{\gamma_0 \delta_0})}^{-1} T_h(\gamma_0,
\delta_0, a^\dagger)\right) \right]
 \cosh\Biggl\{ \sqrt{\gamma_0 \delta_0} \, a^\dagger - \\
 & &  {(\sqrt{\delta_0})}^{-1}
\sqrt{\gamma_0 }  \biggl[ 1   +     z_1 \biggl( 2a^\dagger -
 {(\sqrt{\gamma_0 \delta_0})}^{-1}  T_h(\gamma_0, \delta_0, a^\dagger)
 \biggl)
 \biggr] b \Biggr\}  \,  e^{z a^\dagger}  |0;+\rangle \,  D_0^\ast , \; \; \; \label{sup-sqee-d-z1}\eeqa
where \be T_h(\gamma_0, \delta_0, a^\dagger) = {\left( \cosh(
\sqrt{\gamma_0 \delta_0} \, a^\dagger)\right)}^{-1} \sinh(
\sqrt{\gamma_0 \delta_0} \, a^\dagger ). \ee

In the special case where $z_1 = 0,$ \eqref{sup-sqee-c-z1} and
\eqref{sup-sqee-d-z1}  reduces to \beqa \nonumber |\psi ; -
\rangle &=&
 \cosh\biggl[ \sqrt{\gamma_0 \delta_0} \, a^\dagger - {(\sqrt{\gamma_0})}^{-1} \sqrt{\delta_0 } \, b^\dagger \biggl]
\,  e^{z_0 a^\dagger}  |0;-\rangle \, C_0  \\   &=& -
{(\sqrt{\gamma_0})}^{-1} \sqrt{\delta_0 } \sinh\biggl[
\sqrt{\gamma_0 \delta_0} \, a^\dagger - {(\sqrt{\delta_0})}^{-1}
\sqrt{\gamma_0 } \, b \biggl] \, e^{z_0 a^\dagger}  |0;+\rangle \,
C_0 \label{sup-sqee-c-z0} \eeqa and \beqa \nonumber |\psi ; +
\rangle &=&
 \cosh\biggl[ \sqrt{\gamma_0 \delta_0} \, a^\dagger - {(\sqrt{\delta_0})}^{-1} \sqrt{\gamma_0 } \, b \biggl]
\,  e^{z_0 a^\dagger}  |0;+\rangle \, D_0^\ast  \\   &=& -
{(\sqrt{\delta_0})}^{-1} \sqrt{\gamma_0 } \sinh\biggl[
\sqrt{\gamma_0 \delta_0} \, a^\dagger - {(\sqrt{\gamma_0})}^{-1}
\sqrt{\delta_0 } \, b^\dagger \biggl] \, e^{z_0 a^\dagger} |0;-
\rangle \, D_0^\ast,  \label{sup-sqee-d-z0} \eeqa respectively. By
combining both equations   \eqref{sup-sqee-c-z0} and
\eqref{sup-sqee-d-z0}, we can express the set of  independent
solutions in the form \be  \widetilde{| \psi ; -  \rangle} =
 \exp\left( \sqrt{\gamma_0 \delta_0} \,
a^\dagger - {(\sqrt{\gamma_0})}^{-1} \sqrt{\delta_0 } \, b^\dagger
\right)  \, e^{z_0 a^\dagger}  |0;-\rangle \, {\tilde C}_0
\label{su2c} \ee and \be \widetilde{|\psi ; + \rangle} =
\exp\left( \sqrt{\gamma_0 \delta_0} \, a^\dagger -
{(\sqrt{\delta_0})}^{-1} \sqrt{\gamma_0 } \, b \right) \, e^{z_0
a^\dagger }  |0; + \rangle \, {\tilde D}_0 . \label{su2d} \ee
Thus, we recover  the structure of the spin ${1\over 2}$
representation algebra eigenstates associated to the subalgebra
$\{a, J_+ , J_- \}$ of the $h(2) \oplus su(2)$ Lie algebra
\cite{NaVh}.

\subsection{The general case}
\label{sec-reduction}
 Let us   solve now   the eigenvalue equation
\eqref{gen-super-eq}. The discussion at the end of section
\ref{sec-standar} shows that it can be reduced to a simpler one by
expressing the eigenstate $ |\psi \rangle $ as: \be
\label{squee-red-oper} |\psi \rangle =
 {\bold G} (\beta_0, \gamma_1 , \delta_1 )| \varphi \rangle .\ee Indeed, inserting
\eqref{squee-red-oper} into \eqref{gen-super-eq} and multiplying
by the inverse of the supersqueeze operator $ {\bold G} (\beta_0,
\gamma_1 , \delta_1 ), $ we get \be \label{gen-super-eq-red} [ a +
{\hat \beta}_1 a^\dagger + \gamma_0 b + \delta_0  b^\dagger ]
|\varphi \rangle = z |\varphi \rangle, \ee where \be {\hat \beta
}_1 = \beta_1 + \delta_0 \gamma_1 + \gamma_0 \delta_1 \qquad \in
{\mathbb C}B_{L_1}. \label{betauno}\ee

We can show that, see Appendix \ref{sec-caso-general} section
\ref{sec-cas-gen}, two classes of independent solutions of the
eigenvalue equation \eqref{gen-super-eq-red} exit and are given by
\beqa \nonumber |\varphi ; - \rangle &=& \biggl[ \sum_{\ell \;
{\rm even}}^\infty \exp\left( - {{\hat \beta }_1 {(\gamma_0
\delta_0)}^{-1} \over 2 } \; \ell \right) {\cal O}_{{a^\dagger}}
(\ell, \gamma_0, \delta_0, z_1 ) e^{z a^\dagger} | 0 ; - \rangle
\\ &-& \sum_{\ell \; {\rm odd}}^\infty \exp\left( - {{\hat \beta }_1
{(\gamma_0 \delta_0)}^{-1} \over 2 } \; (\ell - 1) \right)
  {\cal O}_{{a^\dagger}} (\ell, \delta_0, \gamma_0, z_1
) e^{za^\dagger} | 0; + \rangle \biggr] C_0 \label{varphic} \eeqa
and \beqa \nonumber  |\varphi , + \rangle &=& \biggl[ \sum_{\ell
\; {\rm even}}^\infty \exp\left( - {{\hat \beta }_1 {(\gamma_0
\delta_0)}^{-1} \over 2 } \; \ell \right) {\cal O}_{{a^\dagger}}
(\ell, \delta_0, \gamma_0, z_1 ) e^{z a^\dagger} | 0 ; + \rangle
\\ &-& \sum_{\ell \; {\rm odd}}^\infty \exp\left( - {{\hat \beta }_1
{(\gamma_0 \delta_0)}^{-1} \over 2 } \; (\ell - 1) \right)
  {\cal O}_{{a^\dagger}} (\ell, \gamma_0, \delta_0, z_1
) e^{za^\dagger} | 0; - \rangle \biggr] D_0^\ast, \label{varphid}
\eeqa where $C_0$ and $D_0^\ast $ are arbitrary and invertible
Grassmann constants.

Using the results \eqref{calo-g0d0} for the ${\cal
O}_{{a^\dagger}} (\ell, \gamma_0, \delta_0, z_1 )$ operator, we
get \beqa \nonumber |\varphi ; -  \rangle &=& \Biggl[
\cosh(\sqrt{\gamma_0 \delta_0 - {\hat \beta }_1 } \, a^\dagger)
\left( 1 + T_h (\gamma_0, \delta_0, {\hat \beta }_1, a^\dagger )
\sqrt{\gamma_0 \delta_0 - {\hat \beta }_1}
z_1 \right) e^{- z_1 a^\dagger} e^{z a^\dagger} |0;- \rangle \\
&-&
 {(\gamma_0 )}^{-1} \sinh(\sqrt{\gamma_0
\delta_0 - {\hat \beta }_1 } \, a^\dagger) \sqrt{\gamma_0 \delta_0
+ {\hat \beta }_1} e^{- z_1 a^\dagger} e^{z a^\dagger} |0;+
\rangle \Biggr] C_0 \label{sup-sqee-c-b1} \eeqa and \beqa
\nonumber |\varphi ; +  \rangle &=& \Biggl[ \cosh(\sqrt{\gamma_0
\delta_0 - {\hat \beta }_1 } \, a^\dagger) \left( 1 + T_h
(\gamma_0, \delta_0, {\hat \beta }_1, a^\dagger ) \sqrt{\gamma_0
\delta_0 - {\hat \beta }_1}
z_1 \right) e^{- z_1 a^\dagger} e^{z a^\dagger} |0;+ \rangle \\
&-&
 {(\delta_0 )}^{-1} \sinh(\sqrt{\gamma_0
\delta_0 - {\hat \beta }_1 } \, a^\dagger) \sqrt{\gamma_0 \delta_0
+ {\hat \beta }_1} e^{- z_1 a^\dagger} e^{z a^\dagger} |0;-
\rangle \Biggr] D_0^\ast,\label{sup-sqee-d-b1} \eeqa where \be
T_h(\gamma_0, \delta_0, {\hat \beta }_1, a^\dagger) = {\left(
\cosh( \sqrt{\gamma_0 \delta_0 - {\hat \beta }_1 } \,
a^\dagger)\right)}^{-1} \sinh( \sqrt{\gamma_0 \delta_0 - {\hat
\beta }_1} \, a^\dagger ). \ee

\subsubsection{\mathversion{bold}Generalized Spin ${1\over 2}$ representation AES  structure}
 In the special case where $z_1 = 0,$ \eqref{sup-sqee-c-b1} and \eqref{sup-sqee-d-b1}  reduces to
 \beqa \nonumber |\varphi ; -  \rangle &=&
 \exp\left(- {1\over 2} {(\gamma_0)}^{-1} {\hat \beta }_1 a^\dagger \, b^\dagger
 \right)\\ &&
 \cosh \biggl[ \sqrt{\gamma_0 \delta_0 - {\hat \beta }_1 } \, a^\dagger -
 {(\gamma_0)}^{-1} \sqrt{\gamma_0 \delta_0 + {\hat \beta }_1 }  \, b^\dagger \biggl]
\,  e^{z_0 a^\dagger}  |0;-\rangle \, C_0  \label{sup-sqee-cc-b1}
\eeqa and \beqa \nonumber |\varphi ; + \rangle &=&
 \exp\left(- {1\over 2} {(\delta_0)}^{-1} {\hat \beta }_1 a^\dagger \, b  \right)
\\ & & \cosh\biggl[ \sqrt{\gamma_0 \delta_0 - {\hat \beta }_1 } \, a^\dagger -
 {(\delta_0)}^{-1} \sqrt{\gamma_0 \delta_0 + {\hat \beta }_1 }  \, b  \biggl]
\,  e^{z_0 a^\dagger}  |0;+\rangle \, D_0^\ast
\label{sup-sqee-dd-b1} \eeqa respectively. Thus, we get a set of
generalized SAES  that contains the  set of AES associated to the
spin ${1\over 2}$ representation  that we have studied in the
section \ref{sec-spin-unmedio}.

\section{Isospectral harmonic oscillator Hamiltonians  having odd interaction terms}
\label{sec-isospectral} In this section we search for some
isospectral harmonic oscillator systems which are  characterized
by a Hamiltonian admitting an annihilation operator which is  a
Grassmannian linear combination of the generators of the  H-W Lie
superalgebra, i.e., of the form \be {\cal A} = a + \beta a^\dagger
+ \gamma b + \delta b^\dagger, \qquad \beta, \gamma, \delta, \in
{\mathbb C}B_L . \label{calAelemnt} \ee

A family of non-equivalent such Hamiltonians ${\cal H}$ can be
contructed if first we consider a superHermitian Hamiltonian
${\cal H}_0 $ such that the commutator is given by \be
\label{comm-rais-low}[{\cal H}_0 , {\cal A}_0 ] = - {\cal A}_0,
\quad {\rm and} \quad {\cal A}_0 |E_0 ; \pm \rangle =0, \ee where
\be {\cal A}_0 = a + {\hat \beta }_1 a^\dagger + \gamma_0 b +
\delta_0 b^\dagger, \qquad \gamma_0, \delta_0 \in {\mathbb
C}B_{L_0}, \ee  $ {\hat \beta}_1 $ is given by \eqref{betauno} and
$ |E_0 ; \pm \rangle $ are the zero eigenvalue eigenstates of
${\cal H}_0.$ In this way, ${\cal A}_0 $ is effectively  an
annihilation operator and its associated superalgebra eigenstates
a class of supercoherent states for the system characterized by
the Hamiltonian ${\cal H}_0 .$ Second, according to the analysis
of section \ref{sec-cas-gen}, it is possible to construct ${\cal
H}$ satisfying \be [{\cal H} , {\cal A}] = - {\cal A}\ee by taking
\be \label{non-eqivalent}  {\cal A} =  {\bf G} (\beta_0, \gamma_1,
\delta_1 ) {\cal A}_0 {\left( {\bf G} (\beta_0, \gamma_1, \delta_1
)\right)}^{-1} \quad {\rm and } \quad {\cal H} = {\bf G} (\beta_0,
\gamma_1, \delta_1) {\cal H}_0  {\left( {\bf G}( \beta_0 ,
\gamma_1 , \delta_1)\right)}^{-1}, \label{calhghog-1} \ee where $
{\bf G}(\beta_0, \gamma_1, \delta_1 )$ is the standard
supersqueeze operator defined in \eqref{stand-super-sqee}. We see
that our original problem thus reduce to one of finding ${\cal
H}_0 .$  We observe that, the  Hamiltonian ${\cal H}$ in
\eqref{calhghog-1} is not superHermitian  but it belongs to a
class of Hamiltonians  that generalize the one of
$\eta$--pseudo--Hermitian Hamiltonians\cite{AMostafazadeh}.
Indeed, it satisfies the relation \be {\cal H}^\ddagger = {\bf
\eta} {\cal H} {\bf \eta }^{-1}, \ee where $ {\bf \eta}$ is the
superHermitian operator\be {\bf \eta } = {\left( {\bf G}^{-1}
(\beta_0, \gamma_1, \delta_1) \right)}^{\ddagger}{\bf G}^{-1}
(\beta_0, \gamma_1, \delta_1) . \ee  Let us mention that a family
of ${\cal H}_0$--equivalent Hamiltonians can be obtained if we
replace $ {\bold G}(\beta_0, \gamma_1, \delta_1) $ in
\eqref{non-eqivalent} by a suitable  $OSp(2/2)$ superunitary
operator\cite{BuRaRa} \be {\bold U} ( {\cal X}_0 , \Gamma_1 ,
\Delta_1 ) = \exp\left( {\cal X}_0 { (a^\dagger)^2\over 2} - {\cal
X}_0^\ddagger {a^2 \over 2} + \Gamma_1 a^\dagger b^\dagger +
\Gamma_1^\ddagger ab + \Delta_1 a^\dagger b + \Delta_1^\ddagger a
b^\dagger\right), \ee
 where ${\cal X}_0 \in {\mathbb C}B_{L_0}$
and $\Gamma_1, \Delta_1, \in {\mathbb C}B_{L_1} .$

Let us also mention that if we denote $ {\cal A}_0^\ddagger $ the
adjoint of ${\cal A}_0,$ then, the usual commutator leads to
  \beqa \nonumber [{\cal A}_0 , {\cal
A}_0^\ddagger ] &=&  1 - {\hat \beta}_1^\ddagger {\hat \beta}_1
 \{a , a^\dagger \} + (\delta_0^\ddagger
\delta_0 - \gamma_0^\ddagger \gamma_0 ) [b^\dagger , b ] \\ &+& 2
{\hat \beta}_1 \delta_0^\ddagger a^\dagger b - 2 \delta_0 {\hat
\beta}_1^\ddagger a b^\dagger   + 2   {\hat \beta}_1
\gamma_0^\ddagger a^\dagger b^\dagger - 2 \gamma_0 {\hat
\beta}_1^\ddagger ab  \label{calA0com}\eeqa and we notice that,
under the conditions $\gamma_0 = \delta_0 =0$ or ${\hat \beta}_1 =
0,$ the commutator \eqref{calA0com} becomes a diagonal operator in
the Fock vector basis $ \{ | n , \pm\rangle, \ n \in {\mathbb N}\}
.$

\subsection{\boldmath $h(2)$ generalized isospectral oscillator system} Let us here consider the
particular case where $\gamma_0 = \delta_0=0.$ In this case, the
operator ${\cal A}_0 $ takes the simple form \be {\cal A}_0 = a +
{\hat \beta}_1 a^\dagger \ee and the commutator \eqref{calA0com}
writes \be [{\cal A}_0 , {\cal A}_0^\ddagger ] =  1 - {\hat
\beta}_1^\ddagger {\hat \beta}_1
 \{a , a^\dagger \}.
\ee A class of Hamiltonian $ {\cal H}_0 $ satisfying
\eqref{comm-rais-low} is given by \beqa
 \nonumber {\cal H}_0 &=& (1+ {\hat \beta}_1^\ddagger {\hat \beta}_1 ) \left[ {\cal A}_0^\dagger
{\cal A}_0 + {\hat \beta}_1^\ddagger {\hat \beta}_1 {(a^\dagger
)}^2 a^2\right]
\\ &=& a^\dagger a + {\hat \beta}_1 {(a^\dagger )}^2 + {\hat \beta}_1^\ddagger a^2
+ {\hat \beta}_1^\ddagger {\hat \beta}_1 (a^\dagger a + a
a^\dagger) + {\hat \beta}_1^\ddagger {\hat \beta}_1
{(a^\dagger)}^2 a^2. \label{iso-harm-super} \eeqa We notice that
we are in presence of a superHermitian Hamiltonian of the harmonic
oscillator type  with  nilpotent interaction terms which contain
odd contributions. We also notice that, this hamiltonian can be
expressed in the form \be {\cal H}_0 = {{\cal N}\over 2} + { \cal
M}  + {\cal Q}_+ + {\cal Q}_-,\ee where \be {\cal N} = 2 {\hat
\beta}_1^\ddagger {\hat \beta}_1 (a^\dagger a + a a^\dagger),
\quad {\cal Q}_+ = {\hat \beta}_1 {(a^\dagger)}^2, \quad {\cal
Q}_- = {\hat \beta}_1^\ddagger a^2, \quad { \cal M} = a^\dagger a
- {\cal Q}_+ {\cal Q} _-  .\ee The non-zero super-commutation
relations between these operators are given by \be [{\cal M},
{\cal Q}_\pm] = \pm 2 {\cal Q} _\pm, \qquad \{{\cal Q}_+ , {\cal
Q}_- \} = {\cal N}, \ee i.e., they have almost the structure of
$u(1/1)$ superalgebra. Indeed, here ${\cal N}$ is an even
nilpotent operator such that ${\cal N}^{\, 2} =0.$

According to \eqref{comm-rais-low} and \eqref{iso-harm-super}, a
class of superalgebra eigenstates of ${\cal H}_0 $ can be obtained
by applying  $n$ times  ($n=0,1,2,\ldots $)  the raising operator
${\cal A}_0^\dagger $ on the zero eigenvalue eigenstates of ${\cal
A}_0. $ From \eqref{solvarphi}, we deduce that these latter are
given by \be | E_0 ; j \rangle =\left(1 -{1\over 4} {\hat
\beta}_1^\ddagger {\hat \beta}_1\right) \left[|0 ; j \rangle -
{{\hat \beta}_1 \over \sqrt{2}} |2 ; j \rangle\right], \ee where
$j$ corresponds to the set $\{-, +\}.$

Then, as $ {\cal H}_0 | E_0 ; j \rangle = 0,$ the generated energy
 eigenstates are given by
  \be | E_n
; j \rangle \varpropto {({\cal A}_0^\ddagger)}^n | E_0 ; j \rangle
= \left( {(a^\dagger)}^n + {\hat \beta}_1^\ddagger
\sum_{k=0}^{n-1}{(a^\dagger )}^{(n-1-k)} \, a \, {(a^\dagger )}^k
\right) | E_0 ; j \rangle \ee and the corresponding energy
eigenvalues are $E_n^j =n. $ An orthonormalized version of these
states is given by  \beqa \nonumber
  | E_n ; j \rangle &=&  \left( 1 - {1\over 4} {\hat \beta}_1^\ddagger {\hat \beta}_1 (2n+1) \right )
\\ & & \left[
| n ;j \rangle + {{\hat \beta}_1^\ddagger \over 2} \sqrt{n (n-1)}
| n -2 ;j \rangle -
 {{\hat \beta}_1  \over 2} \sqrt{(n+1) (n+2)} | n + 2 ; j \rangle
 \right], \label{eigen-states-n-j}
\eeqa where $n \in {\mathbb N}.$ From \eqref{eigen-states-n-j}, it
is easy to calulate the action of ${\cal A}_0^\ddagger$ and ${\cal
A}_0$ on the $  | E_n ; j \rangle$ eigenstates, we get \be {\cal
A}_0^\ddagger  | E_n ; j \rangle = \left( 1 - {1\over 2} {\hat
\beta}_1^\ddagger {\hat \beta}_1 (n+1) \right ) \sqrt{n+1} |
E_{n+1} ; j \rangle \ee and \be {\cal A}_0 | E_n ; j \rangle =
\left( 1 - {1\over 2} {\hat \beta}_1^\ddagger {\hat \beta}_1 n
\right ) \sqrt{n} | E_{n-1} ; j \rangle. \ee Thus,  the
orthonormalized energy eigenstates $ | E_n ; j \rangle$ can be
written in the standard form \be  | E_n ; j \rangle= \left( 1 +
{1\over 4} {\hat \beta}_1^\ddagger {\hat \beta}_1 n (n+1)\right )
{{({\cal A}_0^\ddagger)}^n \over \sqrt{n!}} \label{asurvide} | E_0
; j \rangle. \ee This is a complete set of states. Indeed, using
\eqref{eigen-states-n-j}, we can demonstrate  the completeness
property \be \sum_{j} \sum_{n=0}^\infty | E_n ; j \rangle \langle
E_n ; j | = I \otimes I = \sum_{j} \sum_{n=0}^\infty | n ; j
\rangle \langle n ; j |. \ee

On the other hand, we can express the
 $| n ; j \rangle$ states in the form \beqa \nonumber | n ; j \rangle &=&  \left(1
- {1 \over 4} {\hat \beta}_1^\ddagger {\hat \beta}_1 (2n+1)
\right) \biggl[ | E_n ; j \rangle    -  j \sqrt{(n+1)(n+2)}  |
E_{n+2} ; j \rangle \, {{\hat \beta}_1 \over 2} \\ &+& j  \sqrt{n
(n-1)}
 | E_{n-2} ;j
\rangle  \, {{\hat \beta}_1^\ddagger \over 2} \biggr]
,\label{nfoncden}
  \eeqa
then,  from \eqref{asurvide} and after some manipulations,  we get
\be |0 ; j \rangle = \left(1 - {1 \over 4} {\hat \beta}_1^\ddagger
{\hat \beta}_1 \right) \exp\left( {{({\cal A}_0^\ddagger)}^2 \over
2} {\hat \beta}_1 \right) | E_0 ; j \rangle. \ee According to
\eqref{solvarphi}, the coherent states associated to a physical
system characterized by the hamiltonian \eqref{iso-harm-super} can
be written as: \beqa \nonumber |\varphi ; j \rangle &=&
\exp\left[- {\hat \beta}_1 { {( a^\dagger )}^2 \over 2 } - {\hat
z}_1 {\hat \beta}_1 { {( a^\dagger )}^3 \over 3
} \right] {\mathbb D} ({\hat z}_0 ) {\mathbb D} ({\hat z}_1 )  \\
& & \left(1 - {1 \over 4} {\hat \beta}_1^\ddagger {\hat \beta}_1
\right) \exp\left( {{({\cal A}_0^\ddagger)}^2 \over 2} {\hat
\beta}_1 \right) | E_0 ; j \rangle \, {\hat C} ({\hat z} , {\hat
\beta}_1 ) . \label{sol+-varphi}  \eeqa

\subsection{Spin \boldmath ${1 \over 2}$ generalized isospectral oscillator system}
In the case where ${\hat \beta}_1 = 0 $ and $\gamma_0^\ddagger
\gamma_0 = \delta_0^\ddagger \delta_0,$ the operator ${\cal A}_0 $
takes the form \be {\cal A}_0 = a + \gamma_0 b  + \delta_0
b^\dagger \ee and the commutator \eqref{calA0com} writes \be
[{\cal A}_0 , {\cal A}_0^\ddagger ] = 1. \label{AAdcan}\ee A class
of Hamiltonian $ {\cal H}_0 $ satisfying \eqref{comm-rais-low} is
given by \be
 \nonumber {\cal H}_0  =    {\cal A}_0^\ddagger
{\cal A}_0  =  a^\dagger a + \gamma_0^\ddagger \gamma_0 + \gamma_0
a^\dagger b + \gamma_0^\dagger a b^\dagger + \delta_0 a^\dagger
b^\dagger + \delta_0^\ddagger a b. \label{iso-harm-spin} \ee We
notice  that this  is a superHermitian Hamiltonian, without
defined parity, which is a linear Grassmann combination of
generators of the $ osp(2/2) \sdir sh(2/2) $ Lie superalgebra.
Then, in  this aspect, the corresponding  Hamiltonian ${\cal H}$
defined in \eqref{non-eqivalent}, complement the classes of
Hamiltonians considered by Buzano et al.\cite{BuRaRa}.

By construction, the eigenstates of ${\cal A}_0 $ corresponding to
the eigenvalue $z=0$ are eigenstates of ${\cal H}_0 $
corresponding to the eigenvalue $E_0 = 0.$ Let us to take these
states to be the normalized version of states
(\ref{su2c}-\ref{su2d}), when $z_0 = 0,$ that is \beqa  |E_0 , -
\rangle &=&  {\left(\sqrt{ 1 + {(\sqrt{\gamma_0})}^{-1}
{({(\sqrt{\gamma_0})}^{-1} )}^\ddagger \sqrt{\delta_0 }
{(\sqrt{\delta_0 })}^\ddagger }\right)}^{-1} \nonumber \\ &&
{\mathbb D } ( \sqrt{\gamma_0 \delta_0} )  \left[|0;- \rangle -
{(\sqrt{\gamma_0})}^{-1} \sqrt{\delta_0 } \, |0;+ \rangle \right ]
\eeqa and
 \beqa  |E_0 , +
\rangle & =& {\left(\sqrt{ 1 + {(\sqrt{\delta_0})}^{-1}
{({(\sqrt{\delta_0})}^{-1} )}^\ddagger \sqrt{\gamma_0 }
{(\sqrt{\gamma_0 })}^\ddagger }\right)}^{-1} \nonumber  \\ &&
{\mathbb D } ( \sqrt{\gamma_0 \delta_0} )    \left[|0;+\rangle -
{(\sqrt{\delta_0})}^{-1} \sqrt{\gamma_0 } \, |0; - \rangle \right
]   . \eeqa Thus, from \eqref{comm-rais-low} and \eqref{AAdcan},
we deduce that a class of orthonormalized eigenstates  of $ {\cal
H}_0 $ corresponding to the eigenvalue $E_n^j = n$ is given by
($n=0,1,2,\ldots; j=-,+$)
 \be |E_n , j \rangle = {{({\cal
A}_0^\ddagger)}^n \over \sqrt{n!}} |E_0 , j  \rangle .\ee
Moreover, a class of normalized coherent states for this
generalized harmonic system which are eigenstates of ${\cal A}_0 $
corresponding to the eigenvalue $z=z_0 $ is easily constructed
as\cite{NaVh} \be |z_0 , j \rangle = \exp\left( z_0 {\cal
A}_0^\ddagger - z_0^\ddagger {\cal A}_0 \right) |E_0 , j
\rangle.\ee These coherent states are obtained from those of
equations (\ref{su2c}-\ref{su2d}) by acting with the following
superunitary transformation \be {\cal U} (z_0; \gamma_0, \delta_0
) = \exp\left[ z_0 (\gamma_0^\ddagger b^\dagger +
\delta_0^\ddagger b) - z_0^\ddagger (\gamma_0 b + \delta_0
b^\dagger) \right]. \ee

\section{Conclusions}
In this paper we have generalized the AES\cite{Brif} concept to
the one of  SAES. We have demonstrate that the SAES associated to
the H--W Lie superalgebra contain the sets of standard coherent
and supercoherent states associated to the usual and
supersymmetric harmonic oscillator systems,
respectively\cite{NaVh,ArZy,Faetal,Pere}. Also, these SAES contain
both the standard squeezed and supersqueezed
states\cite{MMNieto,OrSa} and the supersqueezed states associated
to the spin--${1\over 2}$ representation of the AES  of the
$h(2)\oplus su(2)$ algebra\cite{NaVh}. Let us mention that the
introduction of Grassmann coefficients in the linear combination
of the superalgebra generators helps us to understand  the role
played by the c-numbers (even Grassmann numbers) and d-numbers
(odd Grassmann numbers) interaction coefficients, in the mentioned
literature. Moreover, from the idea of giving to the SAES the
interpretation of an operator associated to a physical system, we
have constructed some classes of superHermitian and ${\bf
\eta}$--pseudo--superHermitian
Hamiltonians\cite{Dewit,AMostafazadeh}, isospectral to the
standard harmonic oscillator hamiltonian.  We have found their
physical eigenstates and their associated supercoherent states. In
this respect, we see that the  SAES concept constitute an
alternative and unified approach for the construction of
generalized coherent and supercoherent and also squeezed and
supersqueezed states for a given quantum system.

\section*{Acknowledgments} The authors' research was
partially supported by research grants from NSERC of Canada and
FQRNT of Qu\'ebec. N.~A.~M. acknowledges financial support from
the ISM.

\appendix

\section{Notations and conventions}\label{sec-Not} In this  appendix we want to fix the
notations and  conventions used in this work. They concern
principally  the concepts of  Grassmann algebra, Lie superalgebra
and their representations, superHermitian and superunitary
operators, super Lie algebra and  linear  Lie supergroup.

Let us remind  that a complex {\bf Grassmann algebra}, ${\mathbb
C} B_L , $ is a linear vector space over the field of complex
numbers, associative   and $Z_2$ graded. It may thus be decomposed
into $ {\mathbb C} B_{L_{0}} + {\mathbb C} B_{L_{1}} , $ where the
even space ${\mathbb C} B_{L_0}$   is generated by the set of
$2^{L-1}$ linearly independent generators ${\cal E}_\mu $ of even
level
 and the odd space ${\mathbb C} B_{L_1} $  is generated by
the set of $2^{L-1}$ linearly independent  generators ${\cal
E}_\mu $ of odd level.   Here, the index  $\mu$ represents either
the empty set $\phi$ or the set $(j_1 ,j_2, \ldots, j_{N(\mu)})$
of $N(\mu)$ integer numbers such that $1 \le j_1 < j_2  \cdots <
j_{N(\mu)} \le L.$ $N(\mu)$ is the level of the generator ${\cal
E}_\mu $. The identity of the algebra is  ${\cal E}_\phi = {\bf
1}$  and ${\cal E}_\mu = {\cal E}_{ {j_1}} {\cal E}_{ {j_2}}
\cdots {\cal E}_{ {j_{N(\mu)}}}$ is the ordered product of $N( \mu
) $ odd generators of level $1$ taken among the set of basic
generators $\{{\cal E}_j, j=1,2, \ldots, L \}. $ The product of
these generators is associative and antisymmetric. Moreover,  any
non zero product of the type $ {\cal E}_{j_1} {\cal E}_{j_2}
\ldots {\cal E}_{j_r}$ of $r$ generators is linearly  independent
of the products containing less than $r$ generators and we have $
{\cal E}_{\phi} {\cal E}_j = {\cal E}_j {\cal E}_{\phi} = {\cal
E}_j, \ \forall j=1,2,\ldots, L.$ The graduation is introduced by
defining the degree of ${\cal E}_{\mu},$ that is
\begin{equation} {\rm deg} \; {\cal E}_{\mu} =
{(-1)}^{N(\mu)},
\end{equation}
with $N(\phi) =0.$

Any  element $B \in {\mathbb C} B_L$  can be written either in the
form
\begin{equation}
B = \sum_{\mu} B_\mu {\cal E}_\mu, \qquad   B_\mu \in {\mathbb C},
\end{equation}
or  as the sum of its even part $B_0 $ and its odd part $B_1$,
i.e., $B= B_0 + B_1$ with
\begin{equation}
B_0 = \sum_{{\rm even} \, N (\mu )} B_\mu {\cal E}_\mu, \qquad B_1
= \sum_{ {\rm odd} \, N(\mu ) \;} B_\mu {\cal E}_\mu.
\end{equation}
We also deduce the graded operations for the Grassmman algebra,
i.e., for all $B_0 , Z_0 \, \in {\mathbb C}B_{L_0}, $ $B_1 , Z_1
\, \in {\mathbb C}B_{L_1}, $ we have \be B_0 Z_0 = Z_0 B_0 \in
{\mathbb C}B_{L_0}, \qquad B_0 Z_1 = Z_1 B_0 \in  {\mathbb
C}B_{L_1}, \qquad B_1 Z_1 = - Z_1 B_1  \in {\mathbb C}B_{L_0}. \ee
In particular,   for all $B=B_0 + B_1 \in {\mathbb C}B_L $ and
$Z_1 \in  {\mathbb C}B_{L_1}$, \be B Z_1 = Z_1 B^\ast, \qquad Z_1
B = B^\ast Z_1, \ee where \be B^\ast = B_0 -B_1, \ee is the {\bf
conjugate} of $B.$  The product of any two elements of the
algebra, $B$ and $B^\prime $, corresponds to
\begin{equation}
BB^\prime = \sum_{\mu} \sum_{\mu^\prime } B_\mu B^\prime
_{\mu^\prime} ({\cal E}_\mu {\cal E}_{\mu^\prime}),
\end{equation}
with \be \label{producto} {\cal E }_\mu {\cal E }_{\mu^\prime } =
\pm {\cal E }_\nu,  \quad {\rm where } \quad  N(\nu) = N(\mu) +
N(\mu^\prime ), \ee when neither of the indices  in the sets
represented by $\mu $ and $\mu^\prime $ are repeated, and ${\cal
E}_\mu {\cal E}_{\mu^\prime} =0, $ when at least one of the index
in the set represented by $\mu$ and $\mu^\prime $ is repeated. The
sign   $\pm$ in  (\ref{producto}) is  determined by using the
antisymmetric property  of the basic generators ${\cal E}_j$ when
reordering the their product.

The identity component of the element $B,$ usually called the
body, is denoted  by $ \epsilon (B) = B_\phi \  \in {\mathbb C},$
whereas the nilpotent quantity $ s (B) = B - B_\phi {\cal
E}_\phi,$ defines the soul of $B.$

With respect to the {\bf complex conjugate} of the element $B \in
{\mathbb C}B_{L},$ we follow the conventions of
Cornwell\cite{Jcor} and thus write \be {\bar B} = \sum_{\mu} {\bar
B}_\mu {\cal E}_\mu, \ee i.e., the basis elements ${\cal E}_\mu $
are considered as the real Grassmann numbers. Also, the {\bf
adjoint} of $B$ is defined by the relation
\begin{equation}
\label{superadjunto} B^{\ddagger} = \sum_{\mu} {\bar B}_\mu {\cal
E}^\ddagger_\mu,
\end{equation}
where
\begin{equation}
{\cal E}^\ddagger_\mu   = \cases{ \; {\cal E}_\mu  , & if \,
$N(\mu)$ is even \cr -i {\cal E}_\mu, & if  $N(\mu)$ is odd. \cr}
\end{equation}
This adjoint operation have the same properties than the ones of
the usual adjoint operation for complex matrices.

The inverse of a Grassmann number $B,$ denoted by ${(B)}^{-1}$ is
defined as \be B {(B)}^{-1} = {(B)}^{-1} B = \epsilon_\phi= 1. \ee
It is important to mention that $B$ is invertible if and only if
$B_\phi \ne 0.$

The integration with respect to an odd  Grassmann variable, must
be considered in the  Berezin sense\cite{Berezin-inte}, i.e., if
$\eta \in {\mathbb C} B_{L_1} $, then
\begin{equation}
\int d\eta =0 , \qquad \int \eta d\eta = 1,
\end{equation}
where the integration is taken over all the domain of definition
of $\eta$.

Let us now recall some useful definitions and properties of Lie
superalgebras, supergroups and associated representations.

\begin{defi}   A $(m/n)$ dimensional complex Lie superalgebra ${\cal L}_s, $
is a complex vector space, $Z_2$ graded with respect to a
generalized Lie product, formed from the direct sum of two
subspaces, the even subspace of dimension $m \ge 0$, which we
denotes by   ${\cal L}_0 , $ and the odd subspace of dimension $n
\ge 0$ ($m+n \ge 1$), which we denotes by  ${\cal L}_1 , $ such
that, for all $ a, b \in {\cal L}_s ,$ there exists a generalized
Lie product (supercommutator) $[a,b]$ with the following
properties:

\newcounter{tres}
\begin{list}
{}{\usecounter{tres}}
\item 1 ) $ [a,b] \in {\cal L}_s $, for all $a,b \in {\cal L}_s $;
\item  2 ) for all $a,b,c  \in {\cal L}_s$ and any complex (real) numbers  $\alpha$ and  $\beta,$
\begin{equation}
[\alpha a + \beta b , c] = \alpha [a,c] + \beta [b,c];
\end{equation}
\item  3 ) if $a$ and $b$ are homogeneous elements  of ${\cal L}_s $ then $[a,b]$ is also a homogeneous
element of ${\cal L}_s $ whose degree is $({\rm deg} \; a + {\rm
deg} \; b)$ mod $2$; that is, $[a,b]$ is odd if either $a$ or $b$
 is odd, but $[a,b]$ is even if  $a$ and $b$ are both even or if
 $a$ and $b$ are both odd;
\item 4 ) for any homogeneous elements $a$ and $b$ of ${\cal L}_s $
\be [b,a] = - {(-1)}^{(deg a)(deb b)} [a,b]; \ee
\item   5)  for any three homogeneous elements  $a,b$ and $c$ of ${\cal L}_s , $ we have the
generalized Jacobi identity:
\begin{equation}
[a,[b,c]] {(-1)}^{({\rm deg} \;  a)({\rm deg} \;  c)} + [b,[c,a]]
{(-1)}^{({\rm deg}\; b)({\rm deg}\; a)} + [a,[b,c]] {(-1)}^{({\rm
deg} \; c)({\rm deg} \; b)} =0.
\end{equation}
\end{list}
\end{defi}
We notice that the even subspace ${\cal L}_0 , $ is an ordinary
complex  Lie algebra whereas the odd subspace, $ {\cal L}_1 , $ is
a carrier space for a representation of a Lie algebra ${\cal L}_0
. $

Just as an ordinary  Lie algebra can, in general,  be represented
by a set of complex matrices  a Lie  superalgebra can also be
represented, in general,  by a set of complex matrices.
Nevertheless, the graded character of a superalgebra implies
certain special conditions for the structure of these matrices.

\begin{defi}  Suppose that for every  $ a \in {\cal L}_s ,$  there
exists a matrix $\Gamma (a) $  from the set of complex matrices
partitioned in the form $ (d_0 / d_1) \times  (d_0 / d_1), $ that
we denotes by  $M(d_0 /d_1; {\mathbb C})$, such that
\newcounter{cuatro}
\begin{list}
{}{\usecounter{cuatro}}
\item 1) for all $a,b \in  {\cal L}_s $  and  $\alpha , \beta $ of the field of  ${\cal L}_s$,
\begin{equation}
\Gamma (\alpha a + \beta b) = \alpha \Gamma (a) + \beta \Gamma
(b);
\end{equation}
\item 2) for all $a,b \in  {\cal L}_s, $
\begin{equation}
\Gamma ( [ a , b ])  =  [\Gamma (a), \Gamma (b)];
\end{equation}
 \item 3) if  $a \in  {\cal L}_0 , $ the even subspace of  $ {\cal L}_s $, then $\Gamma (a) $ a la forme
\begin{equation}
\Gamma (a) = \pmatrix{\Gamma_{00} (a)  & {\bf 0} \cr  {\bf 0} &
\Gamma_{11} (a) \cr },
\end{equation}
where $ \Gamma_{00} (a) $ and  $\Gamma_{11} (a)$ are $d_0 \times
d_0 $ and  $d_1 \times d_1 $ dimensional submatrices
respectively; and if $a \in {\cal L}_1, $  the odd subspace  of $
{\cal L}_s $, then $\Gamma (a) $ has the form
\begin{equation}
\Gamma (a) = \pmatrix{{\bf 0} & \Gamma_{01} (a)   \cr  \Gamma_{10}
(a) & {\bf 0}  \cr },
\end{equation}
where $ \Gamma_{01} (a) $ and $\Gamma_{10} (a)$ are $d_0 \times
d_1 $ and $d_1 \times d_0 $ dimensional submatrices respectively.
Then these matrices  $\Gamma (a)$ are said to form a $(d_0 /
d_1)$--dimensional {\bf graded   representation} of ${\cal L}_s. $
\end{list}
\end{defi}

Let ${\cal L}_s $ be a $(m/ n ) $ dimensional complex   Lie
superalgebra  with even basis elements \linebreak $a_1, a_2,
\ldots, a_m $ and odd basis elements  $a_{m+1}, a_{m+2}, \ldots,
a_{m+n} ,$ represented by the set of matrices $\Gamma (a_k ),
k=1,2, \ldots, m+n. $ To each matrix $\Phi (a_k),$ we can
associate a linear operator $\Phi (a_k)$ acting on the carrier
space $\cal W$ of the representation.  This space  is  a $(d_0 +
d_1) $ inner product vector space expanded by a basis formed by
the set of even vectors $ \{|w_j \rangle\}_{j=0}^{d_0}$ and the
set of odd vectors $\{|w_j \rangle\}_{j=d_0 + 1}^{d_0+ d_1}  $ and
this action is defined by the relation \be \Phi (a_k) | w_j
\rangle = \sum_{i=1}^{d_0+ d_1} {\left(\Gamma (a_k )\right)}_{ij}
| w_i \rangle.  \ee Then ${\cal L}_s $ can also be represented by
set of even operators $ \Phi (a_k)$ ($ k=1,2,\ldots, m$) and the
set of odd operators $\Phi (a_k)$ ($ k=m+1,m+2,\ldots, m+n $),
verifying the same super-commutation relations as the basis
elements $a_k$ $(k=1,2,\ldots, m+n) .$

Let $ \cal X $ to be a polynomial function of the ${\cal L}_s $
superalgebra generators, with complex Grassmannian coefficients.
We say that $ \cal X $ is a {\bf superHermitian}
(anti--superHermitian) operator if $ {\cal X} = {\cal X}^\ddagger
$  ( $ {\cal X} = - {\cal X}^\ddagger $ ). In particular, if $
\cal X $ is   a complex Grassmannian linear combination  of the
${\cal L}_s $ superalgebra generators, i.e., \be {\cal X} =
\sum_{j=1}^{m} C^j \Phi (a_j ) + \sum_{k=1}^{n} D^k \Phi
(a_{m+k}), \ee where $C^j \, \in {\mathbb C}B_L$ ($ j=1,2\ldots ,m
) $ and $ D^k \, \in {\mathbb C}B_L $ ($ k=1,2,\ldots, n)$ then
\be {\cal X}^\ddagger = \sum_{j=1}^{m}
 {(\Phi (a_j ))}^\dagger {(C^j)}^\ddagger  + \sum_{k=1}^{n} {(\Phi
(a_{m+k}))}^\dagger {(D^k )}^\ddagger, \ee where the $\dagger $
symbol is reserved for the usual adjoint operation. We say that a
general $ \cal U $ operator  is {\bf superunitary}  if $ {\cal U }
{\cal U}^\ddagger = {\cal U}^\ddagger {\cal U } = I,$ where $I$ is
the identity operator. In particular, if $ {\cal X}$ is an
anti--superHermitian operator, then ${\cal U} = e^{\cal X}$ is a
superunitary operator.

If for $j=1,2,\ldots , m$ and every element
 ${\cal E}_\mu$ of ${\mathbb C} B_L ,$ we define the even operators
\begin{equation}
\label{superm1} M^j_\mu = {\cal E}_\mu  \Phi (a_j)
\end{equation} and for  $k=1,2, \ldots, n$ and every odd element
${\cal E}_\nu$ of ${\mathbb C} B_L ,$ we define the even operators
\begin{equation}
\label{supern2} N^k_\nu = {\cal E}_\nu  \Phi (a_{m+k} ),
\end{equation}
then the set of  $(m+n) 2^{L-1} $ operators  defined by  the
equation (\ref{superm1}) and (\ref{supern2}) form a  basis of a
$(m+n) 2^{L-1} $  dimensional real  Lie algebra, whose Lie product
is given by the usual commutator induced by the generalized Lie
product of  ${\cal L}_s .$  This real Lie algebra is denoted by
${\cal L}_s ({\mathbb C} B_L )$ and is called a {\bf super Lie
algebra}. A general element $M$ of this super Lie algebra  writes
\begin{equation}
M= \sum_{j=1}^{m} \sum_{ {\rm even} \, \mu } X^j_\mu
   M^j_\mu  + \sum_{k=1}^{n} \sum_{{\rm odd} \; \nu
} \Theta^k_\nu  N^k_\nu ,
\end{equation}
where  $X^j_\mu $ and  $\Theta^k_\nu $ are real parameters. Also
we can write this element in the form
 \begin{equation}
 M= \sum_{j=1}^m  X^j M^j + \sum_{k=1}^n \Theta^k N^k,
 \end{equation}
where $X^j = \sum_{{\rm even} \; \mu  } X^j_\mu  {\cal E}_\mu \,
\in { \mathbb R}B_{L_0}, $ $ \Theta^k = \sum_{{\rm odd} \; \nu }
\Theta^k_\nu {\cal E}_\nu \, \in {\mathbb R} B_{L_1} $ and
\begin{equation}
M^j = {\cal E}_\phi \Phi (a_j),  \qquad  N^k = {\cal E}_\phi \Phi
(a_{m+k}).
\end{equation}

Let us end this Appendix by giving a method of construction of a
linear Lie {\bf supergroup}\cite{Rogers}. If ${\cal L}_s ({\mathbb
C} B_L )$ is a real super Lie algebra whose basis elements are
defined by (\ref{superm1})  and (\ref{supern2}), then every linear
Lie group whose associated real Lie super algebra  is given by
${\cal L}_s ({\mathbb C} B_L )$ is a $(m/n )$ linear Lie
supergroup, which we denote by  ${\cal G}_s ({\mathbb C} B_L) $.
The elements near the identity can be parametrized by
\begin{equation}
{\bf G}({\bf X} ; {\bf \Theta }) = \exp\{M \} = \exp\left\{
\sum_{j=1}^{m} X^j M^j + \sum_{k=1}^{n}   \Theta^k  N^k  \right\}.
\end{equation}

\section{Solving  \boldmath $[a+ \beta a^\dagger+ \gamma b + \delta b^\dagger ] |\psi \rangle = z  |\psi \rangle$}
\label{sec-caso-general}
 In this appendix we will solve  the
eigenvalue equation \eqref{gen-super-eq}.  We will do it in two
steps. Firstly, we will solve the eigenvalue equation
\eqref{super-a-b-bd} and express its solutions in terms of a
generalized supersqueeze operator  acting on the supercoherent
states $e^z | 0; \pm \rangle.$  This supersqueeze operator is used
to reduce the eigenvalue equation \eqref{gen-super-eq} to a
simpler one, see section \ref{sec-reduction}, that is to the
eigenvalue equation \eqref{gen-super-eq-red}. Finally,  we will
solve  the eigenvalue equation \eqref{gen-super-eq-red}.

\subsection{The SAES of \boldmath $ a + \gamma b + \delta b^\dagger$ }
\label{sec-gen-super-squee-oper} Let us solve the eigenvalue
equation \eqref{super-a-b-bd}. The  solution is assumed  on the
type \eqref{sol-type-bos-fer} and by inserting it into
\eqref{super-a-b-bd}, then using the usual properties of the
operators and the states $\{| n ; \pm \rangle \},$ we get the
system ($n = 0,1,2\ldots$)
\beqa \sqrt{n+1} C_{n+1} + \gamma D_n^\ast &=& z C_n, \\
\sqrt{n+1} D_{n+1} + \delta C_n^\ast &=& z D_n. \eeqa Let us
notice the symmetric form of this system. Proceeding by iteration
we can express the $C_n $ and $D_n $ coefficients  in terms of the
arbitrary Grassmann constants $C_0$ and $D_0, $ that is ($n=1,2,
\ldots $)\beqa \nonumber C_n &=& {1 \over \sqrt{n!}} \biggl\{ z^n
C_0 - \sum_{k_1 = 0}^{(n-1)} z^{(n-1- k_1)} \gamma {(z^\ast)}^{k_1
} D_0^\ast  + \sum_{k_1 = 0}^{(n-2)} \sum_{k_2 = 0}^{(n-2-k_1)}
z^{(n-2- k_1
-k_2)} \gamma {(z^\ast )}^{k_2} \delta^\ast z^{k_1}  C_0 \\
\nonumber &-& \sum_{k_1 = 0}^{(n-3)} \sum_{k_2 = 0}^{(n-3-k_1)}
\sum_{k_3 = 0}^{(n-3-k_1-k_2)} z^{(n-3- k_1 -k_2-k_3 )} \gamma
{(z^\ast )}^{k_3 } \delta^\ast z^{k_2} \gamma {( z^\ast )}^{k_1 }
D_0^\ast + \ldots \\ &+& {(-1)}^n  {( \gamma \delta^\ast
)}^{\left[ {n \over 2 }\right]} { \gamma}^{( n - 2 \left[ {n \over
2 }\right])} F_{ n - 2 \left[{n\over 2}\right]} \biggr \},
\label{cn-a-b-bd} \eeqa and \beqa \nonumber D_n &=& {1 \over
\sqrt{n!}} \biggl\{ z^n D_0 - \sum_{k_1 = 0}^{(n-1)} z^{(n-1-
k_1)} \delta {(z^\ast)}^{k_1 } C_0^\ast + \sum_{k_1 = 0}^{(n-2)}
\sum_{k_2 = 0}^{(n-2-k_1)} z^{(n-2- k_1
-k_2)} \delta {(z^\ast )}^{k_2} \gamma^\ast z^{k_1}  D_0 \\
\nonumber &-& \sum_{k_1 = 0}^{(n-3)} \sum_{k_2 = 0}^{(n-3-k_1)}
\sum_{k_3 = 0}^{(n-3-k_1-k_2)} z^{(n-3- k_1 -k_2-k_3 )} \delta
{(z^\ast )}^{k_3 } \gamma^\ast z^{k_2} \delta {( z^\ast )}^{k_1 }
C_0^\ast + \ldots \\ &+& {(-1)}^n  {( \delta \gamma^\ast
)}^{\left[ {n \over 2 }\right]} { \delta}^{( n - 2 \left[ {n \over
2 }\right])} G_{ n - 2 \left[{n\over 2}\right]} \biggr \},
\label{dn-a-b-bd}\eeqa where $\left[{n \over 2}\right]$ represents
the entire part of $ n \over 2  $ and $F_0 = C_0, \; F_1 =
D_0^\ast, \, G_0 = D_0, \, G_1 = C_0^\ast .$ Here we need to
calculate the multiple summation. By expressing $z$ as a sum of
their even and odd parts, $ z = z_0 + z_1, $ we get for example, $
(\ell =1,2,\ldots, n) $ \beqa & & \nonumber \sum_{k_1 = 0}^{(n-
\ell)} \sum_{k_2 = 0}^{(n-\ell- k_1 )} \cdots \sum_{k_{\ell} =
0}^{(n- \ell - k_1 - k_2 - \ldots - k_{\ell -1} )} z^{(n- \ell -
k_1 - k_2 - \ldots - k_{\ell} )} \gamma {(z^\ast )}^{k_{\ell} }
\delta^\ast z^{k_{\ell -1}} \gamma {(z^\ast )}^{k_{\ell - 2}}
\delta^\ast \cdots \\ &=& \nonumber  {n! \over (n-\ell)! \ell !}
\Biggl\{ \stackrel{ \ell \; {\rm factors }}{\overbrace{(\gamma
\delta^\ast \gamma \delta^\ast \cdots )}} z_0  \\  &+& {( n-\ell)
\over \ell + 1 } \sum_{j=0}^{\ell} {(-1)}^{ j + \ell} \stackrel{
(\ell - j ) \; {\rm factors }}{\overbrace{(\gamma \delta^\ast
\gamma \delta^\ast \cdots )}} z_1 \stackrel{  j \; {\rm factors
}}{\overbrace{(\cdots \gamma \delta^\ast \gamma \cdots )}}
 \Biggr\} z_0^{(n-\ell -1 )}\nonumber \\ &=& {\cal O}_{{z_0}}(\ell, \gamma, \delta^\ast, z_1 ) z^n, \label{ozn-op} \eeqa
where  ${\cal O}_{{z_0}} $ is the  differential operator \beqa
\nonumber {\cal O}_{{z_0}} (\ell, \gamma, \delta^\ast, z_1) &=&
{1\over \ell !} \Biggl\{ \stackrel{ \ell \; {\rm factors
}}{\overbrace{(\gamma \delta^\ast \gamma \delta^\ast \cdots )}}
\left( {\partial^\ell \over \partial z_0^{\ell} } - z_1
{\partial^{\ell +1} \over \partial z_0^{\ell +1}} \right) \\
 &+& { 1 \over \ell + 1 } \sum_{j=0}^{\ell} {(-1)}^{ j + \ell}
\stackrel{ (\ell - j ) \; {\rm factors }}{\overbrace{(\gamma
\delta^\ast \gamma \delta^\ast \cdots )}} z_1 \stackrel{  j \;
{\rm factors }}{\overbrace{(\cdots \gamma \delta^\ast \gamma
\cdots )}}   {\partial^{\ell +1} \over
\partial z_0^{\ell+1}}
 \Biggr\} , \label{calo-oper-diff} \eeqa which is also defined for $ \ell=0, $ in fact
 $ {\cal O}_{{z_0}} (0, \gamma, \delta^\ast, z_1 ) =1. $
By inserting \eqref{ozn-op} into \eqref{cn-a-b-bd} and
\eqref{dn-a-b-bd}, we get the compact form of  $C_n$ and $D_n $
coefficients, that is  \be \label{cn-compact} C_n =  \sum_{\ell
=0}^n {(-1)}^{\ell} {\cal O}_{{z_0}} (\ell, \gamma, \delta^\ast,
z_1 ) {z^n \over \sqrt{n!}} F_{ \ell - 2 \left[{\ell \over 2}
\right]} \ee and \be \label{dn-compact} D_n =   \sum_{\ell =0}^n
{(-1)}^{\ell} {\cal O}_{{z_0}} (\ell, \delta, \gamma^\ast, z_1)
{z^n \over \sqrt{n!}} G_{ \ell - 2 \left[{\ell \over 2}
\right]}.\ee By inserting \eqref{cn-compact} and
\eqref{dn-compact} into \eqref{sol-type-bos-fer} and then
separating the terms to multiply arbitrary constants $C_0$ and
$D_0 ,$ we obtain two independent solutions for the eigenvalue
equation \eqref{super-a-b-bd}:  \be \label{sol-c-a-b-bd}| \psi ; -
\rangle = \biggl[ \sum_{n=0}^\infty  \sum_{\ell \; {\rm even}}^{2
[n /2]}
 {\cal O}_{{z_0}} (\ell, \gamma, \delta^\ast, z_1 )
{z^n \over \sqrt{n!}} | n ; - \rangle - \sum_{n=1}^\infty
\sum_{\ell \; {\rm odd}}^{2[(n+1) /2]-1}  {\cal O}_{{z_0}} (\ell,
\delta, \gamma^\ast, z_1 ) {z^n \over \sqrt{n!}} | n ; + \rangle
\biggr] C_0 \ee and \be \label{sol-d-a-b-bd} |\psi ; + \rangle
=\biggl[ \sum_{n=0}^\infty  \sum_{\ell \; {\rm even}}^{2[n /2]}
{\cal O}_{{z_0}} (\ell, \delta, \gamma^\ast, z_1 ) {z^n \over
\sqrt{n!}} | n ; + \rangle - \sum_{n=1}^\infty \sum_{\ell \; {\rm
odd}}^{2[(n+1) /2]-1}   {\cal O}_{{z_0}} (\ell, \gamma,
\delta^\ast, z_1) {z^n \over \sqrt{n!}} | n ; - \rangle \biggr]
D_0^\ast. \ee As $ {\cal O}_{{z_0}} (\ell, \gamma, \delta^\ast,
z_1) z^n = 0,  $ when $\ell > n, $ we can spread out the sum on
$\ell $ index up to infinity and then place it out of the sum
corresponding to the $n$ index. In this way,  we can add up on the
$n$ index and express \eqref{sol-c-a-b-bd} and
\eqref{sol-d-a-b-bd} on the form \be | \psi ; - \rangle = \biggl[
\sum_{\ell \; {\rm even}}^\infty
  {\cal O}_{{z_0}} (\ell, \gamma, \delta^\ast, z_1 )
e^{z a^\dagger} | 0 ; - \rangle - \sum_{\ell \; {\rm odd}}^\infty
  {\cal O}_{{z_0}} (\ell, \delta, \gamma^\ast, z_1 )
e^{za^\dagger} | 0; + \rangle \biggr] C_0 \ee and \be |\psi ; +
\rangle = \biggl[ \sum_{\ell \; {\rm even}}^\infty {\cal
O}_{{z_0}} (\ell, \delta, \gamma^\ast, z_1 ) e^{z a^\dagger} | 0 ;
+ \rangle - \sum_{\ell \; {\rm odd}}^\infty   {\cal O}_{{z_0}}
(\ell, \gamma, \delta^\ast, z_1 ) e^{za^\dagger} | 0 ; - \rangle
\biggr] D_0^\ast, \ee respectively. Finally, using the fact that $
{\partial^\ell \over
\partial z_0^\ell  } e^{z a^\dagger}  = {(a^\dagger)}^\ell e^{z a^\dagger} , $ we
get the generalized supersqueezed states \eqref{sup-sqee-c-a-b-bd}
and \eqref{sup-sqee-d-a-b-bd}.

\subsection{The SAES of \boldmath $a + {\hat \beta}_1 a^\dagger + \gamma_0  b + \delta_0 b^\dagger $ }
\label{sec-cas-gen} Let us solve the eigenvalue equation
\eqref{gen-super-eq-red} by taking   $|\varphi \rangle $ again on
the form \eqref{sol-type-bos-fer}. By inserting it into
\eqref{gen-super-eq-red}, and proceeding as in the above sections,
we get the algebraic system ($n=1,2\ldots $) \beqa \sqrt{n+1}
C_{n+1} +
\gamma_0 D_n^\ast +  \sqrt{n} {\hat {\hat \beta }}_1  C_{n-1} &=& z C_n, \\
\sqrt{n+1} D_{n+1} + \delta_0 C_n^\ast + \sqrt{n} {\hat \beta}_1
  D_{n-1} &=& z D_n, \eeqa
 together with
 \beqa
C_1 =z C_0 - \gamma_0 D_0^\ast, \\
D_1 =z D_0 - \delta_0 C_0^\ast.
 \eeqa
Again, we notice  the symmetric form of this algebraic system.
Proceeding by iteration, we can express the $C_n $ and $D_n $
coefficients  in terms of the arbitrary Grassmann constants $C_0$
and $D_0, $ we get ($n=2,3,\ldots $) \beqa\nonumber  C_n &=&
{\tilde C}_n - {1\over \sqrt{n!}} \Biggl[ \; \; \sum_{{\rm even}
\; \ell =2}^{2 [{n\over 2}]} \sum_{k_1=0}^{(n-\ell )} \sum_{k_2
=0}^{( n-\ell- r_1 )}\sum_{k_3 =0}^{(n-\ell - r_2 )} \cdots
\sum_{k_{\ell-1 } =0}^{(n-\ell - r_{\ell-2} )}
\\ \nonumber  & & \sum_{j=1}^{{\ell \over
2}}( k_{2j-1} +1 ) z^{(n-\ell - r_{\ell - 1})} {(z^\ast)}^{k_{\ell
-1} } z^{k_{\ell - 2}} \cdots {(z^\ast)}^{k_1} {(\sqrt{\gamma_0
\delta_0})}^{\ell - 2} {\hat \beta }_1 \Biggr] C_0 \\ \nonumber
&+& {1\over \sqrt{n!}} \Biggl[ \; \; \sum_{{\rm odd} \; \ell
=3}^{2 [{n+1 \over 2}]-1} \sum_{k_1=0}^{(n-\ell )} \sum_{k_2
=0}^{( n-\ell- r_1 )}\sum_{k_3 =0}^{(n-\ell - r_2 )} \cdots
\sum_{k_{\ell-1 } =0}^{(n-\ell - r_{\ell-2} )}
\\ \label{cnlongue}  & & \sum_{j=1}^{[{\ell \over
2}]}( k_{2j} +1 ) z^{(n-\ell - r_{\ell - 1})} {(z^\ast)}^{k_{\ell
-1} } z^{k_{\ell - 2}} \cdots {z}^{k_1} {(\sqrt{\gamma_0
\delta_0})}^{\ell - 3} \gamma_0 {\hat \beta }_1 \Biggr] D_0^\ast,
\eeqa \beqa\nonumber  D_n &=& {\tilde D}_n - {1\over \sqrt{n!}}
\Biggl[ \; \; \sum_{{\rm even} \; \ell =2}^{2 [{n\over 2}]}
\sum_{k_1=0}^{(n-\ell )} \sum_{k_2 =0}^{( n-\ell- r_1 )}\sum_{k_3
=0}^{(n-\ell - r_2 )} \cdots \sum_{k_{\ell-1 } =0}^{(n-\ell -
r_{\ell-2} )}
\\ \nonumber  & & \sum_{j=1}^{{\ell \over
2}}( k_{2j-1} +1 ) z^{(n-\ell - r_{\ell - 1})} {(z^\ast)}^{k_{\ell
-1} } z^{k_{\ell - 2}} \cdots {(z^\ast)}^{k_1} {(\sqrt{\gamma_0
\delta_0})}^{\ell - 2} {\hat \beta }_1 \Biggr] D_0 \\ \nonumber
&+& {1\over \sqrt{n!}} \Biggl[ \; \; \sum_{{\rm odd} \; \ell
=3}^{2 [{n+1 \over 2}]-1} \sum_{k_1=0}^{(n-\ell )} \sum_{k_2
=0}^{( n-\ell- r_1 )}\sum_{k_3 =0}^{(n-\ell - r_2 )} \cdots
\sum_{k_{\ell-1 } =0}^{(n-\ell - r_{\ell-2} )}
\\ \label{dnlongue}   & & \sum_{j=1}^{[{\ell \over
2}]}( k_{2j} +1 ) z^{(n-\ell - r_{\ell - 1})} {(z^\ast)}^{k_{\ell
-1} } z^{k_{\ell - 2}} \cdots {z}^{k_1} {(\sqrt{\gamma_0
\delta_0})}^{\ell - 3} \delta_0 {\hat \beta }_1 \Biggr] C_0^\ast,
\eeqa where \be r_\ell = \sum_{j=1}^\ell k_j \ee and, in
accordance with Eqs. \eqref{cn-compact} and \eqref{dn-compact},
\be \label{cn-compact-zero} {\tilde C}_n = \sum_{\ell =0}^n
{(-1)}^{\ell} {\cal O}_{{z_0}} (\ell, \gamma_0, \delta_0, z_1 )
{z^n \over \sqrt{n!}} F_{ \ell - 2 \left[{\ell \over 2} \right]}
\ee and \be \label{dn-compact-zero} {\tilde D}_n =   \sum_{\ell
=0}^n {(-1)}^{\ell} {\cal O}_{{z_0}} (\ell, \delta_0 , \gamma_0,
z_1) {z^n \over \sqrt{n!}} G_{ \ell - 2 \left[{\ell \over 2}
\right]}.\ee   Using the fact that for $\ell $ even, we have \be
z^{(n-\ell - r_{\ell - 1})} {(z^\ast)}^{k_{\ell -1} } z^{k_{\ell -
2}} \cdots {(z^\ast)}^{k_1} = z_0^{(n -\ell) } + [(n - \ell) - 2
(k_1 + k_3 + \ldots k_{l-1} ) ]z_0^{(n-\ell-1)} z_1, \ee for $\ell
$ odd, we have  \be z^{(n-\ell - r_{\ell - 1})}
{(z^\ast)}^{k_{\ell -1} } z^{k_{\ell - 2}} \cdots {z}^{k_1} =
z_0^{(n-\ell)} + [ (n- \ell ) - 2 (k_2+k_4+ \ldots + k_{l-1} )
]z_0^{(n-\ell-1)} z_1, \ee and that \be \sum_{k_1=0}^{(n-\ell )}
\sum_{k_2 =0}^{(n-\ell - r_1 )} \sum_{k_2 =0}^{(n-\ell - r_2 )}
\cdots \sum_{k_{\ell-1 } =0}^{(n-\ell - r_{\ell-2} )} \Lambda_\ell
(k) \ee is equal to \be \cases{{(n-1)! \over (n -\ell )! (\ell
-1)!}, & if $\Lambda_\ell (k) = 1 $ \, and \,  $\ell \ge 2,$ \cr
{(n-1)! \over 2 (n -\ell -1)! (\ell -1)!}, & if $\Lambda_\ell (k)
=(k_1+k_3+ \ldots + k_{\ell -1} )$ \cr & and \,  $\ell = 2,4,
\ldots,$ \cr {\ell (n-1)! \over 2 (n -\ell -1)! (\ell + 1)!}
\left[(n - \ell ) + {\ell \over 2} (n -\ell +1 )\right], & if
$\Lambda_\ell (k) ={(k_1+k_3+ \ldots + k_{\ell -1} )}^2$ \cr & and
\, $\ell = 2,4, \ldots,$ \cr {(\ell -1)(n-1)! \over 2 (n -\ell
-1)! \ell ! }, & if $\Lambda_\ell (k) =(k_2+k_4+ \ldots + k_{\ell
-1} )$ \cr & and \, $\ell = 3,5, \ldots,$ \cr {(\ell -1)(n - \ell
+1 )(n-1)! \over 4 (n -\ell -1)! \ell ! }, & if $\Lambda_\ell (k)
={(k_2+k_4+ \ldots + k_{\ell -1} )}^2$ \cr & and \, $\ell = 3,5,
\ldots,$ }
 \ee
and after some manipulations, we can reduce \eqref{cnlongue} and
\eqref{dnlongue} to \beqa \nonumber  C_n &=& {\tilde C}_n - {{\hat
\beta }_1 \over 2 \sqrt{n!}} \Biggl[ \; \; \sum_{{\rm even} \;
\ell =2}^{2 [{n\over 2}]} {n! \over (n -\ell)! (\ell -1)!} \left(
z_0^{(n-\ell)} + {(n -\ell) \over (\ell +1)} z_0^{(n - \ell - 1)}
z_1 \right) {(\sqrt{\gamma_0 \delta_0})}^{\ell-2} \Biggr] C_0 \\
&+& {{\hat \beta }_1 \over 2 \sqrt{n!}} \Biggl[ \; \; \sum_{{\rm
odd} \; \ell =3}^{2 [{n+1 \over 2}]-1}
 { (\ell -1) n! \over (n -\ell)! \ell!}  z_0^{(n-\ell)}
{(\sqrt{\gamma_0 \delta_0})}^{\ell-3} \gamma_0  \Biggr] D_0^\ast
\label{cncourte}\eeqa and \beqa \nonumber D_n &=& {\tilde D}_n -
{{\hat \beta }_1 \over 2 \sqrt{n!}} \Biggl[ \; \; \sum_{{\rm even}
\; \ell =2}^{2 [{n\over 2}]} {n! \over (n -\ell)! (\ell -1)!}
\left( z_0^{(n-\ell)} + {(n -\ell) \over (\ell +1)} z_0^{(n - \ell
- 1)}
z_1 \right) {(\sqrt{\gamma_0 \delta_0})}^{\ell-2} \Biggr] D_0 \\
&+& {{\hat \beta }_1 \over 2 \sqrt{n!}} \Biggl[ \; \; \sum_{{\rm
odd} \; \ell =3}^{2 [{n+1 \over 2}]-1}
 { (\ell -1) n! \over (n -\ell)! \ell!}  z_0^{(n-\ell)}
{(\sqrt{\gamma_0 \delta_0})}^{\ell-3} \delta_0  \Biggr] C_0^\ast,
\label{dncourte}\eeqa respectively. Then, using the fact that \be
 {n! \over (n -\ell)! } z_0^{n-l} = \left( {\partial^\ell \over \partial
 z_0^\ell}  - z_1 {\partial^{\ell+1} \over \partial
 z_0^{\ell+1} }\right) z^n, \qquad {n! \over (n -\ell-1)! } z_0^{n-\ell -
 l}z_1 = z_1{\partial^{\ell+1} \over \partial
 z_0^{\ell+1} } z^n,\ee
we can write \eqref{cncourte} and \eqref{dncourte} in the form
\beqa \nonumber  C_n &=& {\tilde C}_n - {{\hat \beta }_1 \over 2
\sqrt{n!}} \Biggl[ \; \; \sum_{{\rm even} \; \ell =2}^{2 [{n\over
2}]} {1 \over (\ell -1)!} \left( \left( {\partial^\ell \over
\partial
 z_0^\ell}  - z_1 {\partial^{\ell+1} \over \partial
 z_0^{\ell+1} }\right) + {z_1  \over (\ell +1 )} {\partial^{\ell+1} \over \partial
 z_0^{\ell+1} } \right) z^n {(\sqrt{\gamma_0 \delta_0})}^{\ell-2} \Biggr] C_0 \\
&+& {{\hat \beta }_1 \over 2 \sqrt{n!}} \Biggl[ \; \; \sum_{{\rm
odd} \; \ell =3}^{2 [{n+1 \over 2}]-1}
 { (\ell -1) \over  \ell! } \left( {\partial^\ell \over \partial
 z_0^\ell}  - z_1 {\partial^{\ell+1} \over \partial
 z_0^{\ell+1} }\right)
{(\sqrt{\gamma_0 \delta_0})}^{\ell-3} \gamma_0  \Biggr] D_0^\ast
\label{cncourtediff} \eeqa \beqa \nonumber  D_n &=& {\tilde D}_n -
{{\hat \beta }_1 \over 2 \sqrt{n!}} \Biggl[ \; \; \sum_{{\rm even}
\; \ell =2}^{2 [{n\over 2}]} {1 \over (\ell -1)!} \left( \left(
{\partial^\ell \over
\partial
 z_0^\ell}  - z_1 {\partial^{\ell+1} \over \partial
 z_0^{\ell+1} }\right) + {z_1  \over (\ell +1 )} {\partial^{\ell+1} \over \partial
 z_0^{\ell+1} } \right) z^n {(\sqrt{\gamma_0 \delta_0})}^{\ell-2} \Biggr] D_0 \\
&+& {{\hat \beta }_1 \over 2 \sqrt{n!}} \Biggl[ \; \; \sum_{{\rm
odd} \; \ell =3}^{2 [{n+1 \over 2}]-1}
 { (\ell -1) \over  \ell! } \left( {\partial^\ell \over \partial
 z_0^\ell}  - z_1 {\partial^{\ell+1} \over \partial
 z_0^{\ell+1} }\right)
{(\sqrt{\gamma_0 \delta_0})}^{\ell-3} \delta_0  \Biggr] C_0^\ast
\label{dncourtediff}, \eeqa respectively. We notice that, when the
inverse of the product $ \gamma_0 \delta_0 $ exist, or even if it
does not exist, we can write formally  these last equations  in
the compact form \beqa \nonumber  C_n &=& {\tilde C}_n - {{\hat
\beta }_1 {(\gamma_0 \delta_0)}^{-1} \over 2 } \Biggl[ \; \;
\sum_{{\rm even} \; \ell =2}^{2 [{n\over 2}]}  \ell \,
   {\cal O}_{z_0} (\ell, \gamma_0,\delta_0,
z_1 ) {z^n \over \sqrt{n!}}\Biggr] C_0 \\ &+&
 {{\hat \beta }_1 {(\gamma_0
\delta_0)}^{-1} \over 2 } \Biggl[ \; \; \sum_{{\rm odd} \; \ell
=3}^{2 [{n+1 \over 2}]-1} (\ell -1) \, {\cal O}_{z_0} (\ell,
\gamma_0,\delta_0, z_1 )  {z^n \over \sqrt{n!}} \Biggr] D_0^\ast
\label{cntildecoeff} \eeqa and \beqa \nonumber  D_n &=& {\tilde
D}_n - {{\hat \beta }_1 {(\gamma_0 \delta_0)}^{-1} \over 2 }
\Biggl[ \; \; \sum_{{\rm even} \; \ell =2}^{2 [{n\over 2}]}  \ell
\,
   {\cal O}_{z_0} (\ell, \delta_0,\gamma_0,
z_1 ) {z^n \over \sqrt{n!}}\Biggr] D_0 \\ &+&
 {{\hat \beta }_1 {(\gamma_0
\delta_0)}^{-1} \over 2 } \Biggl[ \; \; \sum_{{\rm odd} \; \ell
=3}^{2 [{n+1 \over 2}]-1} (\ell -1) \, {\cal O}_{z_0} (\ell,
\delta_0,\gamma_0, z_1 )  {z^n \over \sqrt{n!}} \Biggr] C_0^\ast.
\label{dntildecoeff} \eeqa  Now, by inserting \eqref{cntildecoeff}
and \eqref{dntildecoeff} into \eqref{sol-type-bos-fer} and
proceeding exactly as in section \ref{sec-gen-super-squee-oper}, we
get the two independent solutions \eqref{varphic} and
\eqref{varphid}.



\begin{thebibliography}{alpha}
\bibitem{Nam} Alvarez-Moraga, N. (2000). \emph{\'Etats coh\'erents et
comprim\'es}, MSc. Thesis, Universit\'e de Montr\'eal.
\bibitem{NaVh}Alvarez-Moraga, N., and  Hussin, V. (2002). \emph{Generalized coherent and squeezed states based on the $h(1) \oplus su(2)$ algebra}, \emph{Journal of Mathematical  physics} {\bf 43}, 2063.
\bibitem {ArZy} Aragone, C.,  and Zypman, F. (1986).
\emph{Supercoherent states}, \emph{Journal of  Physics  A: Mathematical and General} {\bf
19}, 2267.
\bibitem {Ba}  Bacry, H. (1978). \emph{Eigenstates of complex linear combinations of $J1,J2,J3$ for any representation of $SU(2)$}, \emph{Journal of Mathematical Physics} {\bf 19}, 1192.
\bibitem{Berezin-inte} Berezin, F. A. (1987). \emph{Introduction to Superanalysis}, Reidel, Dordrecht, the Netherlands
\bibitem{YBeVh}  B\'erub\'e-Lauzi\`ere, Y.,  and Hussin, V. (1993). \emph{Comments of the definitions of coherent states
for the SUSY harmonic oscillator}, \emph{Journal of  Physics A:
Mathematical and General} {\bf 26}, 6271.
\bibitem{Brif0} Brif, C.  (1996). \emph{Two--photon algebra eigenstates: a unified approach to squeezing}, \emph{e--print quant-ph/9605006}.
\bibitem {Brif} Brif, C. (1997). \emph{$SU(2)$ and $SU(1,1)$ algebra eigenstates: A unified analytic approach to coherent and intelligent states}, \emph{International Journal of  Theoretical  Physics} {\bf 36},
1651.
\bibitem{BuRaRa} Buzano, C.,  Rasetti,  M. G.,  and  Rastello, M. L. (1989). \emph{Dynamical Superalgebra of the Dressed Jaynes--Cummings Model}, \emph{Physical Review Letters} {\bf
62}, 137.
\bibitem{JwAsSaVh}Cooper, F.,  and  Freedman, B. (1983). \emph{Aspects of supersymmetric quantum mechanics}, \emph{ Annals
of  Physics} (New York) {\bf 146}, 262.
\bibitem{Jcor}  Cornwell, J. F. (1989). \emph{Group Theory in Physiscs, Vol. III : Supersymmetries and Infinite-dimensional Algebras },
Academic, New York.
\bibitem{Dewit} DeWitt, B. S. (1984). \emph{Supermanifolds}, Cambridge University Press,
Cambridge. UK.
\bibitem{AmElGraLMMi}  El Gradechi, A. M., and Nieto, L. M. (1996). \emph{Supercoherent states, supe K\"ahler geometry and geometric quantization}, \emph{Communications in Mathematical Physics} {\bf 175},
521.
\bibitem{Faetal}  Fatyga, B. W.,  Kosteleck\'y, V. A., Nieto, M. M.,
and Truax, D. R. (1991).  \emph{Supercoherent states},
\emph{Physical Review D} {\bf 43}, 1403.
\bibitem{HuLMNi} Hussin, V.,   and  Nieto, L. M. (1993). \emph{Supergroups factorizations through matrix realization}, \emph{ Journal of Mathematical  Physics} {\bf 34}, 4199.
\bibitem{KoNiTr} Kosteleck\'y, V. A.,   Nieto,  M. M., and  Truax, D. R. (1993). \emph{Supersqueezed States},  \emph{ Physical Review
A} {\bf  48}, 1045.
\bibitem{AMostafazadeh} Mostafazadeh, A. (2002). \emph{Pseudo-Hermiticity versus PT symmetry: The necessary condition for the reality of the spectrum of a non-Hermitian Hamiltonian
}, \emph{Journal of  Mathematical  Physics} {\bf 43}, 205.
\bibitem{MMNieto} Nieto, M. M. (1992). \emph{Coherent states and squeezed states, supercoherent states and supersqueezed states}, \emph{ Proceedings of  Second International Workshop on Squeezed
States and Uncertainty Relations}, Moscow, Russia.
\bibitem{OrSa} Orszag M.,   and Salamo, S. (1988). \emph{Squeezing and minimum uncertainty states
in the supersymmetric harmonic oscillator}, \emph{Journal of
Physics  A: Mathematical and General} {\bf 21}, L1059.
\bibitem{ApCt} Pelizzola, A.  and Topi, C. (1992). \emph{Generalized coherent states for dynamical superalgebras}, \emph{ e-print cond--mat/9209022}.
\bibitem{Pere} Perelomov, A. M. (1986). \emph{Generalized Coherent States and their Applications},
Springer--Verlag, Berlin.
\bibitem{Rogers} Rogers, A. (1981). \emph{Super Lie groups: global topology and local structure},  \emph{Journal of mathematical
Physics} {\bf 22}, 939.
\bibitem{Salam} Salam, A., and Strathdee, J. (1975). \emph{Superfields and Fermi--Bose symmetry}, \emph{Physical Review D} {\bf 11}, 1521.
\bibitem{Salo} Salomonson, P., and Van Holten J. W. (1982). \emph{Fermionic coordinates and supersymmetry in quantum mechanics}, \emph{Nuclear
Physics B} {\bf 196}, 509.
\bibitem{Sa} Sarkar, S. (1991). \emph{Supercoherent states for the $t$--$J$ model}, \emph{Journal of  Physics  A: Mathematical and General}  {\bf 24}, 5775.
\bibitem{WeZu} Wess, J., and Zumino, B. (1974). \emph{Supergauge transformations in four dimensions}, \emph{Nuclear Physics B} {\bf
70}, 39.


\end{thebibliography}
\end{document}